\documentclass[structabstract]{aa}

\usepackage[varg]{txfonts}
\usepackage{graphicx}
\usepackage{epstopdf}
\usepackage{lscape}
\usepackage{natbib}
\bibpunct{(}{)}{;}{a}{}{,} 

\begin{document}
\title{Irradiation dose affects the composition of organic refractory materials in space:} 
\subtitle{results from laboratory analogues}
\titlerunning{Irradiation dose affects the composition of organic refractory materials in space}
\authorrunning{Urso et al.}

\author{
R. G. Urso \inst{1}
\and V. Vuitton \inst{2}
\and G. Danger \inst{3}
\and L. Le Sergeant d'Hendecourt \inst{3}
\and L. Flandinet \inst{2}
\and Z. Djouadi\inst{1}
\and O. Mivumbi \inst{1}
\and F. R. Orthous-Daunay \inst{2}
\and A. Ruf \inst{3}
\and V. Vinogradoff \inst{3}
\and C. Wolters \inst{2}
\and R. Brunetto \inst{1}
}

\institute{ Universit\'{e} Paris-Saclay, CNRS, Institut d'astrophysique spatiale, 91405, Orsay, France\\
\email{rurso@ias.u-psud.fr}
\and Universit\'{e} Grenoble Alpes, CNRS, IPAG, Grenoble F-38000, France
\and Aix-Marseille Universit\'{e}, Laboratoire de Physique des Interactions Ioniques et Moléculaires (PIIM) UMR-CNRS 7345, F-13397 Marseille, France}

\abstract
{Near- and mid-infrared observations have revealed the presence of organic refractory materials in the solar system, in cometary nuclei and on the surface of centaurs, Kuiper-belt and trans-neptunian objects. In these astrophysical environments, organic materials can be formed because of the interaction of frozen volatile compounds with cosmic rays, stellar/solar particles, and favoured by thermal processing. The analysis of laboratory analogues of such materials gives information on their properties, complementary to observations.} 
{We present new experiments to contribute in the understanding of the chemical composition of organic refractory materials in space.} 
{We bombard frozen water, methanol and ammonia mixtures with 40 keV H$^+$ and we warm the by-products up to 300~K. The experiments allow the production of organic residues that we analyse by means of infrared spectroscopy and by Very High Resolution Mass Spectrometry to study their chemical composition and their high molecular diversity, including the presence of hexamethylenetetramine and its derivatives.}
{We find that the accumulated irradiation dose plays a role in determining the residue's composition.} 
{Based on the laboratory doses, we estimate the astrophysical timescales to be short enough to induce an efficient formation of organic refractory materials at the surface of icy bodies in the outer solar system.}


\maketitle

\section{Introduction}
\label{intro}
Astronomical observations allowed the detection of various frozen compounds on the surface of dust grains (icy grain mantles) in the interstellar medium \citep[e.g.][]{vandehulst49, tielenshagen82, whittet96, caselliceccarelli12, boogert15} as well as on the surface of small bodies in the solar system, such as comets, centaurs, and Kuiper-belt objects \citep[e.g.][]{cruikshank98, barucci06, biver06, altwegg17b, stern19}. 
In the early solar system, these bodies formed thanks to the accretion of the material present in the presolar cloud, including icy grain mantles in the outer regions. 
Surviving the sun-formation process, these objects could have preserved, at least partially, information on the composition of the presolar cloud \citep[e.g.][]{biver06, willacy15, altwegg17b, nesvorny19, mckinnon20}.

During their lifetime, frozen volatiles experience both irradiation by UV photons, cosmic-rays (CR) and stellar or solar particles, as well as heating. 
Such processes determine changes in the physical and chemical properties of ices \citep[e.g.][]{tielensallamandola87, greenberg95, ehrenfreund99, urso19}.
Frozen small bodies in the outer solar system are exposed to CR, solar wind (SW) and solar energetic particles (SEP) \citep[e.g.][]{johnson90, cooper03, strazzulla03, urso20}. Such bodies exhibit red slopes in the visible and near-infrared (IR) spectra that are related to the presence of a refractory C-rich material, whose formation is attributed to the irradiation of volatiles on their surfaces \citep[e.g.][]{cruikshank98, barucci06, brown11, grundy20} or to the incorporation of red materials present in the presolar cloud \citep[e.g.][]{dalleore11}.

Laboratory experiments shed light on the effects induced by irradiation and heating on solid-phase matter. 
The irradiation with UV photons or energetic particles (ions or electrons) of frozen volatiles determines the break of molecular bonds and the formation of radicals and molecular fragments that then recombine to form new compounds \citep[e.g.][and references therein for details on the physico-chemical process]{oberg16, rothard17}. 
The warm-up of processed frozen mixtures induces not only the sublimation of volatile compounds \citep[e.g.][]{collings04, aboumrad16, aboumrad17} but also an increase in the molecular diffusion and reactivity \citep{mispelaer13, theule13}. 
As a consequence, the chemical complexity further increases, the initially flat and bright spectra of frozen volatiles show a reddening in the visible and near-IR spectra \citep[][]{brunetto06, poston18} and a refractory material, named organic refractory residue, is eventually formed.  

Organic refractory residues are thought to be representative of materials in comets and at the surface of red frozen bodies in the outer solar system \citep[e.g.][]{agarwal85, strazzullajohnson91, munozcaro03, danger13, baratta15, fresneau17, urso17, accolla18, baratta19}. 
Laboratory IR spectra provide relevant information on the composition of organic refractory residues, revealing the presence of various functional groups \citep[e.g.][]{palumbo04, munozcaro04, vinogradoff13, urso17, demarcellus17}.  
The analysis through mass spectrometry (MS) provides a more detailed characterization of the composition of such samples. MS reveals a high extent of molecular diversity of residues produced after UV photolysis and subsequent warm-up of frozen volatiles, with the detection of thousands of molecules up to 4000 Da \citep[][]{danger13, fresneau17, gautier20} and the presence of prebiotic compounds, such as amino acids \citep{bernstein02, nuevo08}, sugars, sugar derivatives \citep{meinert16, nuevo18}, and nucleobases \citep[e.g.][]{nuevo09, nuevo12, materese13, materese17}.

The chemistry induced by UV photons shows similarities with that induced by energetic ions, even if with different by-products formation cross sections \citep[e.g.][]{baratta02, munozcaro14, rothard17}. One of the main differences arises from the fact that whereas the UV photons penetration is strongly affected by the ice optical constants, ions can penetrate up to meters according to their energy 
and independently from the target optical constants \citep[e.g.][]{cooper03, strazzulla03, urso20}.
Previous work pointed out the existence of another difference among the two types of irradiation, i.e. the production of hexamethylenetetramine (C$_6$H$_{12}$N$_4$, hereafter HMT), a precursor of compounds of prebiotic interest \citep[e.g.][]{hulett71, bernstein95, vinogradoff18, vinogradoff20} and HMT derivatives, i.e. HMT in which a side group substitutes a peripheral H atom.

On the one hand, in UV residues HMT is detected through both IR spectroscopy and mass spectrometry \citep[e.g.][]{bernstein95, munozcaro04, vinogradoff13, danger13}, up to 50\% of the residue mass \citep[e.g.][]{bernstein95, munozcaro03}.
Various works focus on the production of HMT \citep[e.g.][]{bernstein95, woon01, munozcaro04, vinogradoff13, zeffiro16, materese20}.
\citet{vinogradoff12} studied the formation of HMT after UV irradiation and warmup of H$_2$O:CH$_3$OH:NH$_3$ mixtures. In the mechanism they proposed, H$_2$CO and HCOOH are formed at 25~K, with HCOOH acting as a catalyst throughout the reaction. During the warm-up, H$_2$CO reacts with NH$_3$ forming aminomethanol, precursor of methylenimine. At higher temperature, methyleneimine polymerizes as protonated trimethylenetriamine (TMT), and the further warm-up to 300~K determines TMT cyclisation to form HMT. 
A similar mechanism would form HMT derivatives, with side groups added to intermediate compounds \citep{materese20}.

On the other hand, IR spectra of residues produced after ion irradiation of frozen volatiles do not show evidences of HMT and its derivatives. The missing detection of HMT in such samples could be attributed to the non-formation of its precursors after ion bombardment.
Taking into account the results obtained by \citet{baratta94} after ion-bombardment of CH$_3$OH-rich ices, \citet{bernstein95} proposed that ion irradiation favours the conversion of methanol to acetone rather than to H$_2$CO. Thus, the presence of HMT has been proposed as a probe of UV photolysis of astrophysical methanol-rich ices.
However, \citet{hudsonmoore00} have shown that H$_2$CO is formed after ion irradiation of frozen volatiles containing methanol, and GC-MS analysis revealed the presence of HMT in residues produced after ion bombardment of H$_2$O:CH$_3$OH:CO:NH$_3$ mixtures with 800 keV H$^+$ of up to a dose of 25 eV~molecule$^{-1}$ \citep{cottin01}.

In this work, we use in-situ IR spectroscopy to characterize H$_2$O:CH$_3$OH:NH$_3$ mixtures deposited at 15~K, bombarded with 40 keV H$^+$ and warmed up to room temperature in order to produce organic refractory residues. 
Such mixtures and the source of processing aim to simulate, at our best, the irradiation of frozen surfaces in the outer solar system by Solar Energetic Particles (SEP), that have a primary role in the processing of such surfaces \citep{urso20}, as well as the heating that surfaces experience during their lifetime.
We use Very High Resolution Mass Spectrometry (VHRMS) to characterize the composition of residues and to point out eventual variations in the chemical composition that are related to the irradiation dose given to the original frozen mixtures. This information is crucial because in space, the dose accumulated by a frozen surface is linked to the timescale of its exposure to energetic particles. Thus, to understand how irradiation dose affects the composition of organic refractory materials can shed light on the timescale of irradiation of an outer body surface. 
Finally, we report on dedicated analysis performed by means of tandem mass spectrometry (MS/HRMS) to reveal the presence of HMT and its derivatives in organic refractory residues, and we investigate on the effects of irradiation dose in the production of such compounds.

\section{Experimental methods} \label{experimentalmethods}
Experiments were performed with the IrradiatioN de Glaces et M\'{e}t\'{e}orites Analys\'{e}es par R\'{e}flectance VIS-IR \citep[INGMAR][]{lantz17} setup at IAS-IJCLab (Orsay, France). 
Water, methanol and ammonia gaseous mixtures are prepared in a mixing chamber (P$\leq$10$^{-4}$ mbar) and are then injected in the main vacuum chamber (P$\sim$5$\times$10$^{-8}$~mbar) that hosts a ZnSe substrate in thermal contact with the cold finger of a closed-cycle He cryocooler (CTI, 14-300~K). A 5~cm long copper tube placed on the backside of the sample holder prevents the deposition on the other side of the substrate.
After deposition, frozen mixtures are bombarded with 40 keV H$^+$ produced by the SIDONIE ion accelerator \citep[IJCLab, Orsay,][]{chauvin04}. 
The ion beam arrives on samples at an angle of 10$^{\circ}$ with respect to the surface normal. The rastering of the beam ensures a homogeneous covering of the sample surfaces. 
During bombardment we integrate the ion current to estimate the fluence (ions cm$^{-2}$). The ion current density is kept below 800~nA~cm$^{-2}$ to avoid the macroscopic heating of the samples.
We then multiply the fluence by the stopping power in eV~cm$^2$/16~u calculated with the SRIM software \citep{ziegler08} to obtain the dose, i.e. the energy deposited per unit volume in the sample by incident radiation, in eV/16~u where u is the unified atomic mass unit \citep[e.g.][]{strazzullajohnson91}.

In our experiments, we produce three residues by bombarding frozen H$_2$O, CH$_3$OH, and NH$_3$ mixtures, varying the irradiation dose.
In order to ensure a uniform irradiation of the whole film thickness, we deposit frozen films whose thickness (about 400~nm) is lower than the penetration depth of 40 keV H$^+$ \footnote{According to calculations with the SRIM software \citep{ziegler08}, 40 keV H$^+$ can penetrate up to 800~nm in a H$_2$O:CH$_3$OH:NH$_3$=1:1:1 mixture assuming a density of 0.8 g~cm$^{-3}$, and H implantation is not negligible above 600~nm.}. To produce thick residues, we perform a multi-deposition and subsequent irradiation procedure \citep{urso20}. After the deposition and irradiation of the first film, we deposit a second film of the same composition that is in turn irradiated up to the same dose of the first film. For each sample, this procedure is repeated three to four times. 
After irradiation, samples are warmed up to room temperature with a constant heating rate of 3~K~min$^{-1}$. 
Organic refractory residues are obtained using the mixture and irradiation doses reported in Table~\ref{experiments}.

\begin{table*}
	\centering
	\caption{\label{experiments} Frozen mixture ratio and irradiation dose of 40 keV H$^+$ used to produce the samples analyzed in this work.}
	\begin{tabular}{cccc}
		\hline
		\hline
		Mixture at 15 K& Dose& Sample& Analysis\\
		H$_2$O:CH$_3$OH:NH$_3$& (eV/16~u)& name\\
		\hline
		1:0:0&32& blank& FT-IR, VHRMS\\
		1:1:0&85&1:1:0+85 eV/16~u& FT-IR\\
		1:1:1&29&1:1:1+29 eV/16~u& FT-IR, VHRMS\\
		3:1:1&67&1:1:1+67 eV/16~u& FT-IR, VHRMS\\
		1:1:1&98&1:1:1+98 eV/16~u& FT-IR, VHRMS\\

		\hline
	\end{tabular}
\end{table*}

Throughout the experiment we perform in-situ Fourier-Transform IR spectroscopy. The IR beam from the internal source of the spectrometer is directed to the vacuum chamber thanks to mirrors on an optical bench, and the IR beam enters the chamber through a ZnSe window. 
The incident beam arrives on the sample with an angle of 10$^{\circ}$ with respect to the surface normal and is then collected by an MCT detector placed on the other side of the vacuum chamber. 
IR spectra are acquired with a resolution of 1~cm$^{-1}$. 
Further information about the experimental setup are given in \citet{lantz17} and \citet{urso20}. 
The residue on the ZnSe substrate is then removed from the HV chamber and stored in a dedicated stainless steel sample holder under static vacuum. 
Prior to the VHRMS characterization, two residues, the 1:1:1+29 eV/16~u and the 3:1:1+67 eV/16~u are stored for 14 and 21 days, respectively, while the 1:1:1+98 eV/16~u is stored for 240 days.

VHRMS is performed with a linear trap/orbitrap mass spectrometer (LTQ Orbitrap XL, ThermoFisher) at Institut de Plan\'{e}tologie et d'Astrophysique de Grenoble (IPAG, Grenoble, France).
Residues are recovered by rinsing five times with 50~$\mu$L of ultrapure methanol. We then collect 50~$\mu$L from the resulting solution and further dilute it in 250~$\mu$L of ultra-pure methanol. 
In order to avoid any degradation of the samples, VHRMS is performed immediately after the collection of each residue from the substrate.
After the rinsing, we analyse the ZnSe substrates by means of FT-IR, to verify whether all residues are efficiently collected. In one case only, i.e. the 1:1:1+98 eV/16~u residue, we observe very weak contributions at about 3200 and 1650~cm$^{-1}$. However, the area of the 3200~cm$^{-1}$ feature is found to be only about the 3\% of the area of the same feature in the residue. This minor contribution could be due to insoluble compounds or to a minor quantity of solution left on the substrate during the residue collection, because of the fast evaporation of methanol.
The solution is injected in the electrospray ionization source (ESI) by means of a Hamilton gastight syringe with a 
rate of a 3~$\mu$L~min$^{-1}$.
ESI has the advantage to limit the fragmentation of molecules, allowing their detection as [M+H]$^{+}$ in positive ionization mode or [M-H]$^{-}$ in negative ionization mode. 
In the Orbitrap, charged species are accumulated in a quadrupolar ion trap and are then transferred into the Orbitrap analyzer. Here, ions oscillate between two side electrodes and around a central electrode and produce a periodic signal that is converted into a mass spectrum using Fourier transform. 
For each sample, we acquired data in three m/z ranges: 50-300, 150-400 and 350-950 m/z, with a resolving power $M/\Delta M$=10$^5$ at m/z~400.
The instrument parameters are listed in the following: spray voltage of 3.5~kV, capillary temperature of 275~$^{\circ}$C, capillary voltage 35~V, 
tube lens voltage 50 and 70~V and -50 and -70~V in positive and negative ionization modes, respectively. 
Spectra are acquired with 4 scans, each consisting of 128 micro-scans \citep[for details on the choice of the number of scans and micro-scans, see][]{wolters20}.
Prior to the samples analysis, the instrument is calibrated with a mixture of L-methionine-arginyl-phenylalanyl-alanine, caffeine and ultramark 1621 for positive ESI mode, sodium dodecyl sulfate, sodium taurocholate and ultramark 1621 for negative ESI mode. 
The data reduction and analysis are based on the method reported by \citet{danger13, danger16} and \citet{fresneau17}, and it allows to get rid of contamination and to attribute each m/z peak to a stoichiometric formula C$_c$H$_h$N$_n$O$_o$.
The Orbitrap resolution and accuracy allows to precisely assign stoichiometric formulas up to m/z=400. 
We also analyse a ZnSe substrate on which we deposited pure water ice that was then bombarded with 40 keV H$^+$ at 15~K (blank experiment). 
	The ion irradiation of pure water determines the formation of hydrogen peroxide \citep[e.g.][]{moorehudson00, urso18}, that sublimates, together with water, during the further warm-up to room temperature. Due to the absence of methanol and ammonia, the eventual detection of C- and N-bearing compounds would be due to contamination on the substrate or in the experimental setup. 
	Thus, all the molecular ions detected in this sample are removed from the m/z spectra of residues, and they are not taken into account for the analysis.

Each molecular ion in the range 50-400 m/z is attributed to the parent molecule by adding a proton in the analysis in negative ESI and by subtracting a proton in the analysis in positive ESI mode. 
	After attributions, we also calculate elemental abundances as unweighted averages, given by ${\sum x}/{\sum C}$ where \textit{x} is the number of nitrogen, oxygen or hydrogen and \textit {C} is the number of carbon for each stoichiometric formula, and weighted averages, calculated as ${\sum x \times I}/{\sum C \times I}$ where \textit{x} is the number of nitrogen, oxygen or hydrogen, \textit{C} is the number of carbon and \textit{I} is the intensity of the related m/z peak. 
In fact, even if the intensity of m/z peaks is affected by variations of the ionization yield \citep[e.g.][]{danger16}, due to the fact that the intensity is proportional to the concentration of the related species \citep{hockaday09}, weighted averages give an estimation of the elemental abundances in samples. 
To have information about the structure of compounds in samples, we calculate the Double Bond Equivalent (DBE), i.e. the degree of unsaturation within the sample, though the formula $DBE=C-\frac{H}{2}+\frac{N}{2}+1$, where C, H and N are the number of carbon, hydrogen and nitrogen atoms in each formula. DBE are then weighted to the intensity of the m/z peak. 

Tandem mass spectrometry (MS/HRMS) is also performed on the residue solutions, using helium as the activation gas in the LTQ ion trap and final mass analysis being performed in the Orbitrap. The normalized collision energy is set to 20\% (the absolute collisional energy is not known) with	an activation time of 30~ms and an isolation window of 1~Da.

In this work, we use van Krevelen plots \citep{vankrevelen50} to obtain information on the composition of our samples. van Krevelen plots are atomic ratio plots that allow a screening of samples with respect to chemical families using H/C, O/C and N/C ratios. Specific zones in the diagrams are related to specific chemical functions and structures, and they have been used to characterize organic matter in meteorites \citep[e.g.][]{schmittkopplin10} as well as organic refractory residues obtained after UV and ion irradiation of frozen volatiles \citep[e.g.][]{gautier14, danger16, ruf19, gautier20}.

\section{Results}
\subsection{Infrared spectroscopy}
\begin{figure*}
	\centering
	\includegraphics[width=1\linewidth]{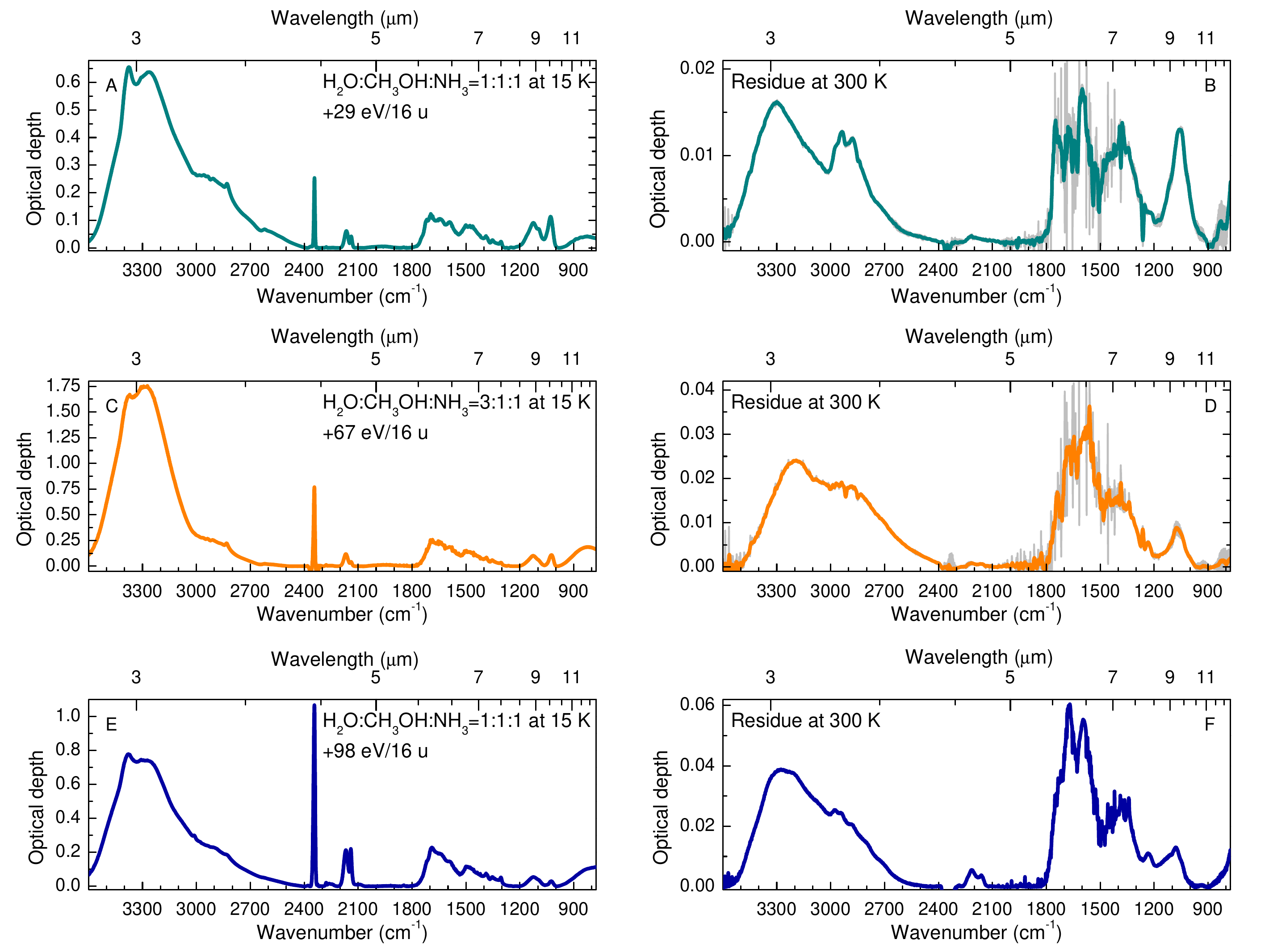}
	\caption{Infrared spectra acquired after irradiation of H$_2$O:CH$_3$OH:NH$_3$ mixtures with 40 keV H$^+$ at 15~K (left panels) and after warm-up to 300~K (right panels). Panel A and B: H$_2$O:CH$_3$OH:NH$_3$=1:1:1 + 29 eV/16 u; panel C and D: H$_2$O:CH$_3$OH:NH$_3$=3:1:1 + 67 eV/16; panel E and F: H$_2$O:CH$_3$OH:NH$_3$=1:1:1 + 98 eV/16 u. Spectra in panels B and D are obtained after the smoothing of the raw spectra (grey lines).}
	\label{fig.Figure_ir}
\end{figure*}

\begin{figure}
	\centering
	\includegraphics[width=1.05\linewidth]{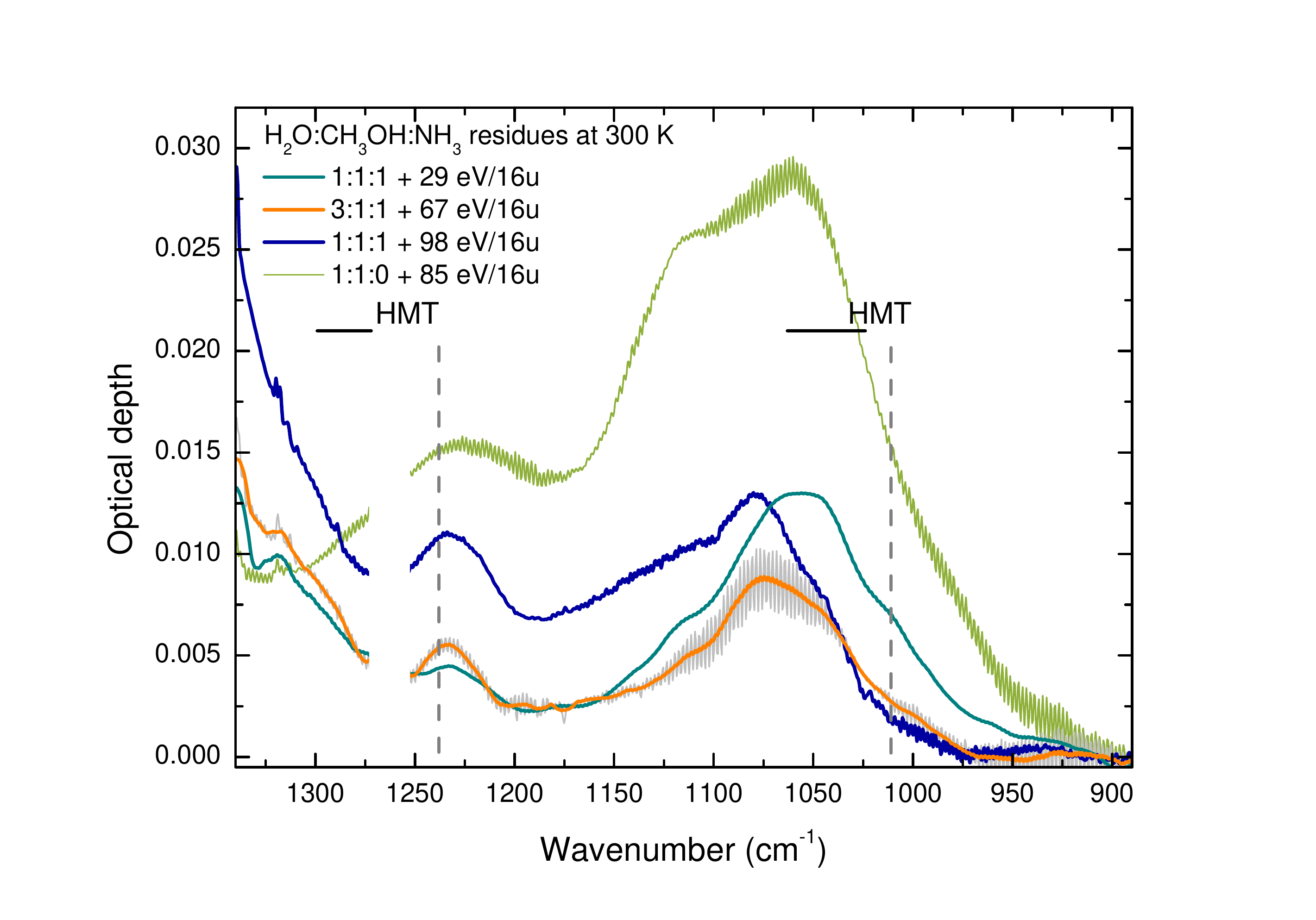}
	\caption{IR spectra in the range 1300-900~cm$^{-1}$ of residues acquired after 30 minutes at 300~K. The spectra of the 1:1:1+29 eV/16~u and of the 3:1:1+67 eV/16~u residues are obtained after the smoothing of the acquired spectra (grey lines in the background). For comparison, we show the spectrum of a 1:1:0+85.2 eV/16~u residue. We do not show spectra between 1253 and 1274~cm$^{-1}$ because of the presence of spikes due to not well compensated water vapour.
		Vertical dashed lines show the positions of pure HMT bands reported by \citet{bernstein95}. Black horizontal lines show the range of HMT and HMT-derivatives bands according to ab-initio calculations by \citet{bera19} and \citet{materese20}. }
	\label{fig:figureres1300-900}
\end{figure}

The IR spectra acquired during the experiments are shown in Fig.~\ref{fig.Figure_ir}, where left panels show the spectra of the first deposited films after the irradiation with 40 keV H$^+$ at 15~K.
In Table~\ref{ice_irradiation} we list the main bands observed in the three samples, the associated vibrational mode and the possible carriers identified in similar experiments reported in literature. 
In the spectra we identify CO$_2$ (2342~cm$^{-1}$) and CO (2140~cm$^{-1}$), whose intensities are found to be higher with increasing the dose given to samples. Their production, as well as that of CH$_4$, H$_2$CO, and C$_2$H$_6$ that are also detected in the spectra, is due to the recombination of unstable by-products obtained after the destruction of CH$_3$OH. 
Also N-bearing species are formed, and their production involves the destruction of NH$_3$ and the formation of NH$_2^{\bullet}$, such as in the case of OCN$^-$, NH$_4^+$, HNCO and NH$_2$HCO \citep{chen07, jheeta13}. 
The presence of HCOO$^-$ is attributed to the reaction between formic acid and ammonia, thus HCOO$^-$ could be an indirect evidence of the presence of HCOOH \citep{vinogradoff11}, that in turn forms thanks to the recombination between HCO and OH radicals \citep{hudsonmoore99}.
Spectra also reveal the presence of H$_2$O, CH$_3$OH and NH$_3$, that are not completely destroyed after irradiation. 

During the warm-up of irradiated samples we observe evident changes in the spectra, and at room temperature organic refractory residues are left on the substrates. After the arrival at 300~K, we note that spectra show slight variations, implying that residues undergo further modifications. 
We thus analyse the residues spectra acquired after about 30 minutes at 300~K (Fig.~\ref{fig.Figure_ir} panels B, D and F).
Due to the low thickness of residues, the signal-to-noise ratio is low in the spectra of the 1:1:1+29 eV/16~u and of the 3:1:1+67 eV/16~u residues. We thus perform a smoothing procedure to allow a better visualization of the spectra. The position of the main bands and the associated vibrational modes are reported in Table~\ref{residues}.
In all residues, N-H, O-H and C-H stretching modes are observed between 3500 and 2500~cm$^{-1}$ and several bands are present between 1500 and 1250~cm$^{-1}$.
The analysis of the spectra reveals the presence of bands associated to various chemical functions. Among them, alcohols and carboxilic acids ($\sim$3300~cm$^{-1}$), nitriles ($\sim$2215~cm$^{-1}$), esters (1740-1720~cm$^{-1}$), amides ($\sim$1680~cm$^{-1}$), alkenes, imines, and aromatics ($\sim$1590~cm$^{-1}$). 
NH$_4^+$ signatures could be present between 2900 and 2600 cm$^{-1}$. NH$_4^+$ could be the counter-ion of CN$^-$, OCN$^-$ whose features are detected at about 2200~cm$^{-1}$, as well as of HCOO$^-$.
Even if such salts would desorb at lower temperature, their presence at 300~K could be due to the trapping in the residue complex matrix, as observed for other volatile species \citep[e.g.][]{urso17}.
At lower wavenumbers, residues show bands peaked at 1234-1231~cm$^{-1}$ and 1078-1056~cm$^{-1}$, due to CO and CN stretching modes. These bands are not symmetric, implying the presence of various subcomponents. The spectra between 1300 and 900 cm$^{-1}$ are shown in Fig.~\ref{fig:figureres1300-900}. In the Figure, we show the position of the CN stretching and asymmetric CN stretching mode bands of pure HMT in Argon matrix given by \citet{bernstein95} and the range of HMT and HMT derivatives bands calculated by \citet{materese20} and \citet{bera19} (black horizontal lines).
The 1:1:1+29 eV/16 u residue clearly shows a shoulder at about 1010~cm$^{-1}$, where pure HMT exhibits its most intense band.   
Even if such bands could be an evidence of HMT in our residues, we note the presence of bands in the same wavelength range in a NH$_3$-free residue produced after irradiation of a H$_2$O:CH$_3$OH=1:1 mixture with 40 keV H$^+$ up to a final dose of 85 eV/16~u (green spectrum in Fig.~\ref{fig:figureres1300-900}).
Ammonia, or at least a source of nitrogen, is needed to produce HMT \citep{bernstein95, vinogradoff12}, thus it cannot be present in such residue. The presence of these bands in the nitrogen-free residue hinders any conclusion on the spectroscopic evidence of HMT in our samples.

\subsection{Very-High Resolution Mass Spectrometry}\label{VHRMS}

\begin{figure}
	\centering
	\includegraphics[width=0.9\linewidth]{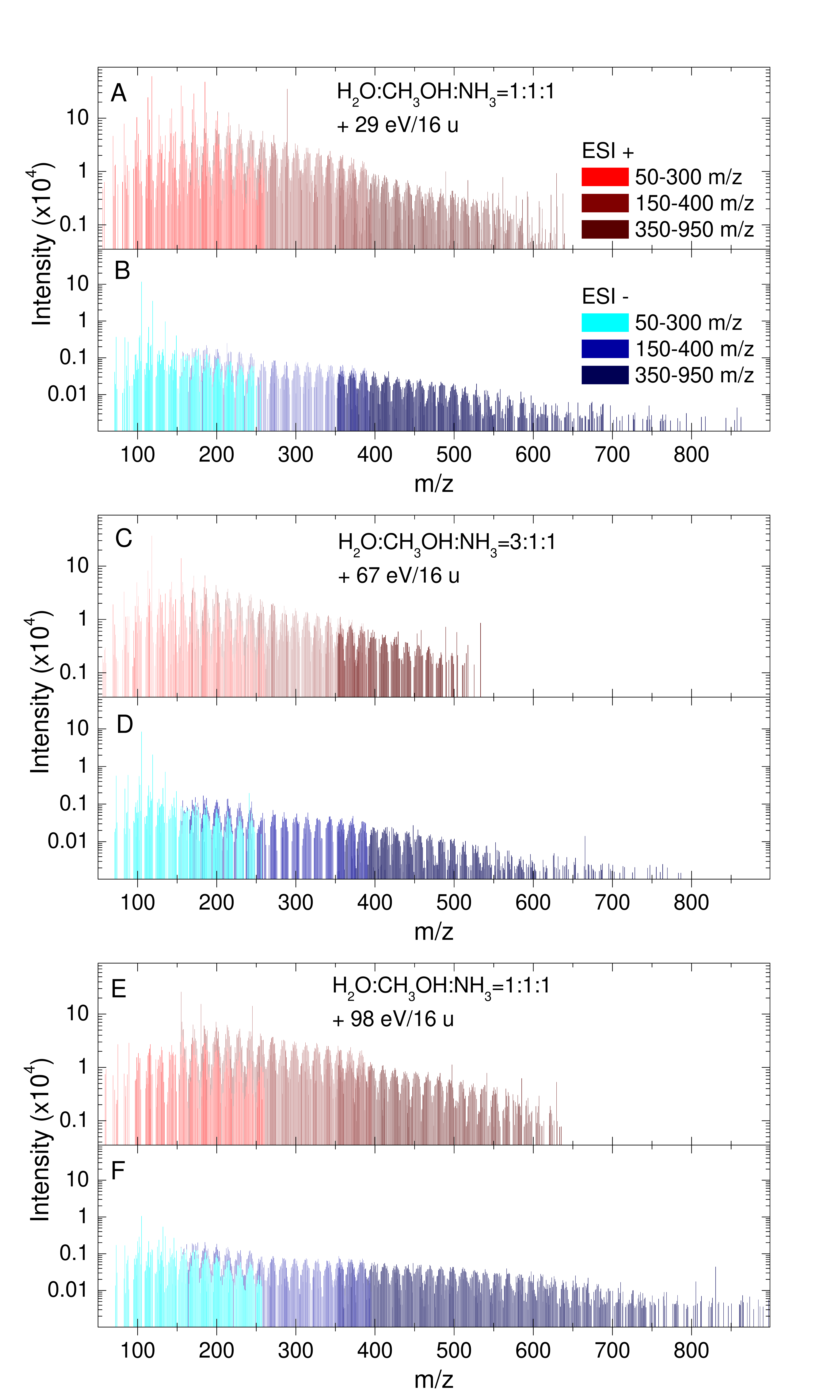}
	\caption{Mass spectra in positive (top panels) and negative (bottom panels) ESI modes of residues synthesised after ion bombardment and warm up of H$_2$O:CH$_3$OH:NH$_3$ frozen mixtures. Panels A and B: 1:1:1 + 29 eV/16~u residue. Panel C and D: 3:1:1 + 67 eV/16~u residue. Panel E and F: 1:1:1 + 98 eV/16~u. }
	\label{fig.Figure_Orbitrap}
\end{figure}

In Fig.~\ref{fig.Figure_Orbitrap} we show the mass spectra of the three residues in the range 50-950 m/z in both positive and negative ESI modes. 
Each spectrum is a composite spectrum obtained by merging spectra acquired in three m/z ranges, 50-300, 150-400 and 350-950 m/z. In the figure, each colour is associated to the m/z range and to the ESI ionization mode.
In positive ESI modes (top panels in Fig.~\ref{fig.Figure_Orbitrap}) peaks are observed up to 550-650 m/z. In negative ESI modes (bottom panels in Fig.~\ref{fig.Figure_Orbitrap}) peaks are present in the whole investigated range, but the signals beyond 600-700 m/z could be attributed to instrumental noise.

\begin{table*}
	\centering
	\caption{\label{elemental_abu} Elemental abundance ratios in organic refractory residues produced after ion bombardment and warm-up of H$_2$O:CH$_3$OH:NH$_3$ mixtures analysed in both positive (p) and negative (n) ESI modes. We estimate Relative Standard Deviation of 6\% for N/C, 12\% for O/C, 3\% for H/C.}
	\begin{tabular}{ccccccccc}
		\hline \hline
		\multicolumn{9}{c}{Range 50-300 m/z}\\
&&&&&&&&\\
		Sample& ESI mode& &Unweighted average&  & &&Weighted average& \\
		\hline
		&         & N/C & O/C & H/C&      & N/C & O/C & H/C \\
		\hline
		1:1:1		&p&0.34&0.30&1.73&&0.38&0.27&1.81\\
		+29 eV/16~u	&n&0.24&0.60&1.59&&0.14&0.87&1.73\\
		&average  	&0.29 &0.45 &1.66 &&0.26&0.57&1.77\\
		\hline
		3:1:1				&p&0.39&0.29&1.74&&0.41&0.29&1.78\\
		+67 eV/16~u	&n&0.30&0.61&1.57&&0.19&0.88&1.71\\
		&average  	&0.35 &0.45 &1.66&&0.30&0.58&1.74\\
		\hline
		1:1:1				&p&0.41&0.29&1.71&&0.43&0.29&1.75\\
		+98 eV/16~u		&n&0.32&0.57&1.47&&0.30&0.67&1.51\\
		&average 	&0.36 &0.43 &1.59 &&0.37&0.48&1.63\\
		\hline
		&&&&&&&&\\
		\multicolumn{9}{c}{Range 150-400 m/z}\\
		&&&&&&&&\\
		Sample& ESI mode& &Unweighted average&  & &&Weighted average& \\
		\hline
		&         & N/C & O/C & H/C&      & N/C & O/C & H/C \\
		\hline
		1:1:1				&p&0.31&0.31&1.70&&0.33&0.31&1.74\\
		+29 eV/16~u	&n&0.27&0.50&1.63&&0.27&0.53&1.64\\
		&average  	&0.29&0.41 &1.67 &&0.30&0.42&1.69\\
		\hline
		3:1:1				&p&0.37&0.30&1.69&&0.38&0.29&1.71\\
		+67 eV/16~u	&n&0.35&0.49&1.62&&0.35&0.51&1.62\\
		&average  	&0.36 &0.40 &1.65&&0.36&0.40&1.66\\
		\hline
		1:1:1				&p&0.38&0.30&1.62&&0.39&0.29&1.65\\
		+98 eV/16~u		&n&0.35&0.48&1.48&&0.35&0.50&1.49\\
		&average 	&0.36 &0.39 &1.55 &&0.37&0.40&1.57\\				
		\hline
	\end{tabular}
\end{table*}

\begin{figure*}
	\centering
	\includegraphics[width=1.05\linewidth]{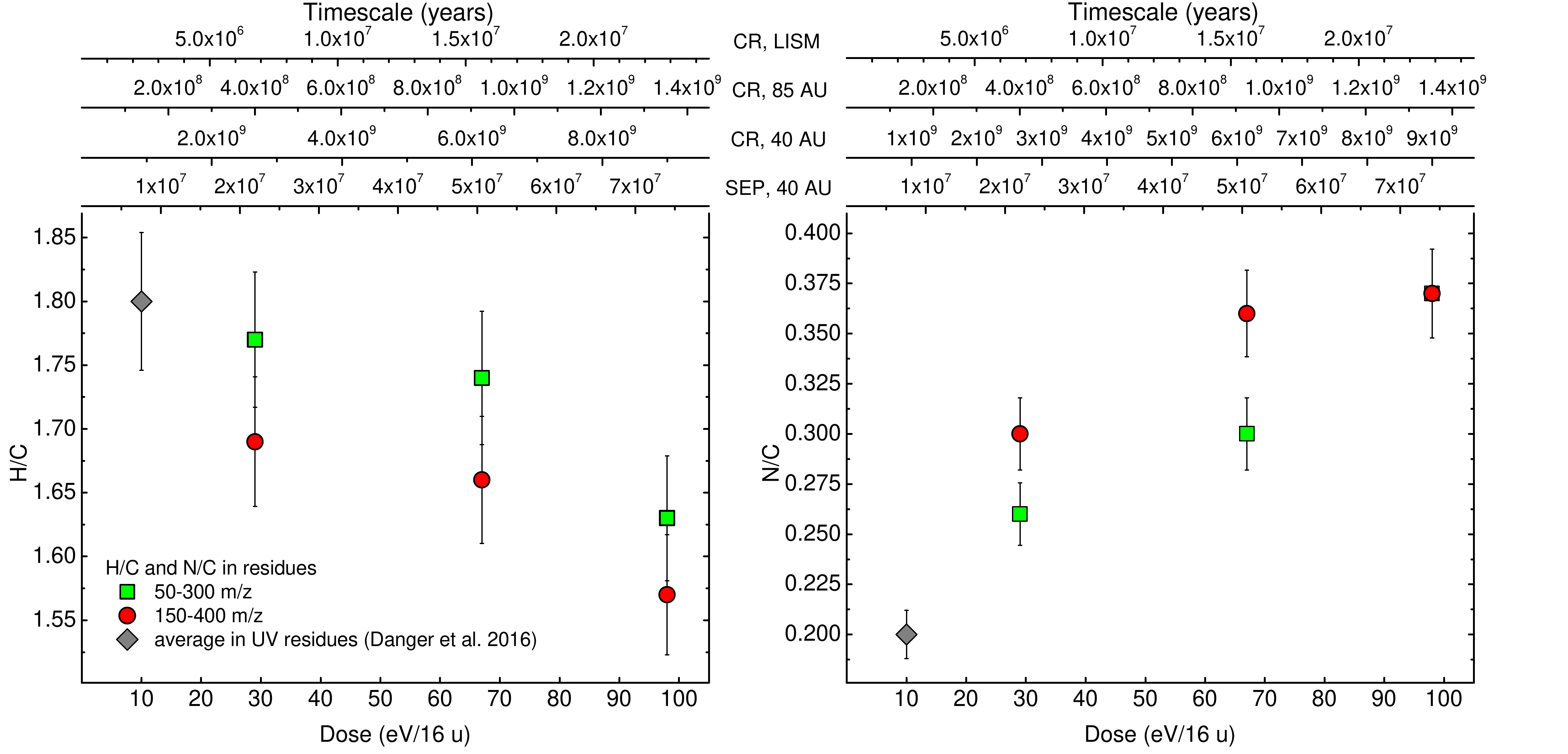}
	\caption{H/C and N/C ratios in organic refractory residues produced after irradiation with 40 keV~H$^+$. Gray rhombus give the average elemental ratios of UV residues analysed by means of VHRMS in both positive and negative electron spry ESI modes in the range 50-400 m/z by \citet{danger16}. The bars refer to the relative standard deviation (RSD) given by \citet{fresneau17}, i.e. 3\% for H/C and 6\% for N/C values. 
	Top x-axes give the timescale necessary for a frozen surface (50~$\mu$m thick) to accumulate the dose given in the bottom x-axes. To calculate timescales, we take into account four cases: bombardment by Solar Energetic Particles (SEP) on a surface located at 40 AU; bombardment by cosmic rays (CR) for a surface at 40 AU; bombardment by CR on a surface at 85 AU, bombardment by CR on a surface in the Local Interstellar Medium (LISM), i.e. beyond 100 AU. Timescales do not take into account the time needed for a surface to experience the heating necessary for the production of organic refractory materials. 
}
	\label{fig:figureelemabuvsdose}
\end{figure*}

We calculate the H, C, O and N elemental abundances in each residue. 
Table~\ref{elemental_abu} reports the N/C, O/C and H/C measured in both positive and negative ESI modes and their average values. 
We observe that N/C ratios increase as the irradiation dose given to the frozen mixtures increases, while H/C ratios follow the opposite behaviour.
The O/C ratio does not show a well defined trend, even if its values are lower in the sample produced with the highest irradiation dose (1:1:1+98 eV/16 u).
It is interesting to note that in UV residues produced after irradiation up to about 10 eV/16 u, using the average of both positive and negative ESI modes in the range 50-400 m/z (analogously to our analysis), \citet{danger16} calculate H/C=1.8 and N/C=0.2. The higher H/C and lower N/C ratios with respect to our samples are thus explained by the lower irradiation dose used for their production.
In Fig.~\ref{fig:figureelemabuvsdose} we plot the H/C and N/C ratios with respect to the dose given to the frozen mixtures, including also the values given by \citet{danger16} for UV residues.
For both UV residues and our samples, the same warm-up rate has been used, i.e. 3~K~min$^-1$.

\begin{figure}
	\centering
	\includegraphics[width=0.7\linewidth]{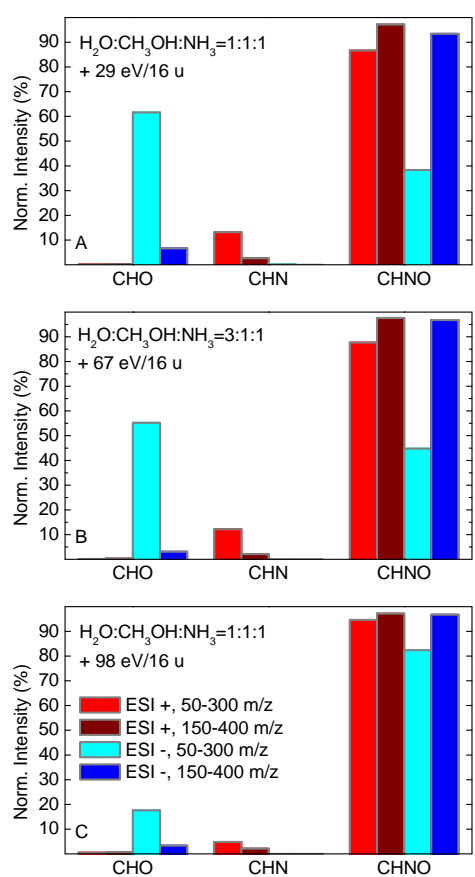}
	\caption{Weighted abundance of CHO, CHN and CHNO molecular groups in organic refractory residues. Panel A: 1:1:1+29 eV/16~u residue; panel B: 3:1:1+67 eV/16 u residue; panel C: 1:1:1+98 eV/16~u residue. Red and dark-red columns refer to the analysis in positive ESI mode in the 50-300 and 150-400 m/z, respectively, while light blue and blue refer to the analysis in negative ESI mode in the 50-300 and 150-400 m/z range, respectively.}
	\label{fig.Figure_moleculargroups}
\end{figure}

\begin{figure*}
	\centering
	\includegraphics[width=0.99\linewidth]{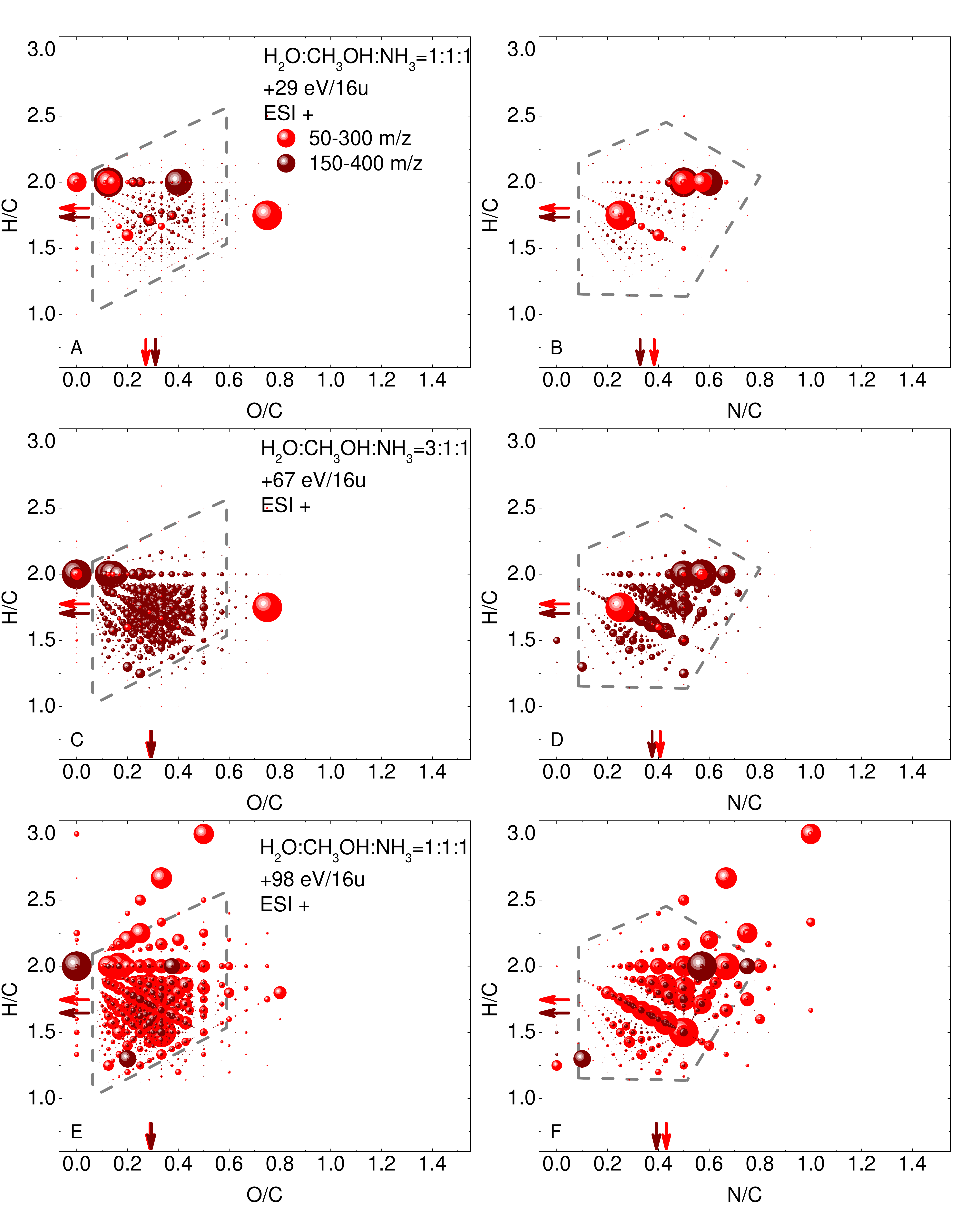}
	\caption{van Krevelen scatter plots of hydrogen to carbon versus oxygen to carbon (left panels) and versus nitrogen to carbon (right panels) in the three residues analysed in positive ESI mode in the 50-300 m/z range (red dots) and 150-400 m/z range (dark-red dots). Panels A and B: 1:1:1+29 eV/16~u residue; panels C and D: 3:1:1+67 eV/16~u residue; panels E and F: 1:1:1+98 eV/16~u residue. The size of the dots is given by the normalized intensity. Gray-dashed lines delimit the zone of highest density of ions observed in the residues analysed by \citet{danger16}. Arrows show the H/C, O/C and N/C weighted averages reported in Table~\ref{elemental_abu} for the range 50-300 m/z (red arrows) and 150-400 m/z (dark-red arrows).}
	\label{fig:figurehcocp}
\end{figure*}

\begin{figure*}
	\centering
	\includegraphics[width=0.99\linewidth]{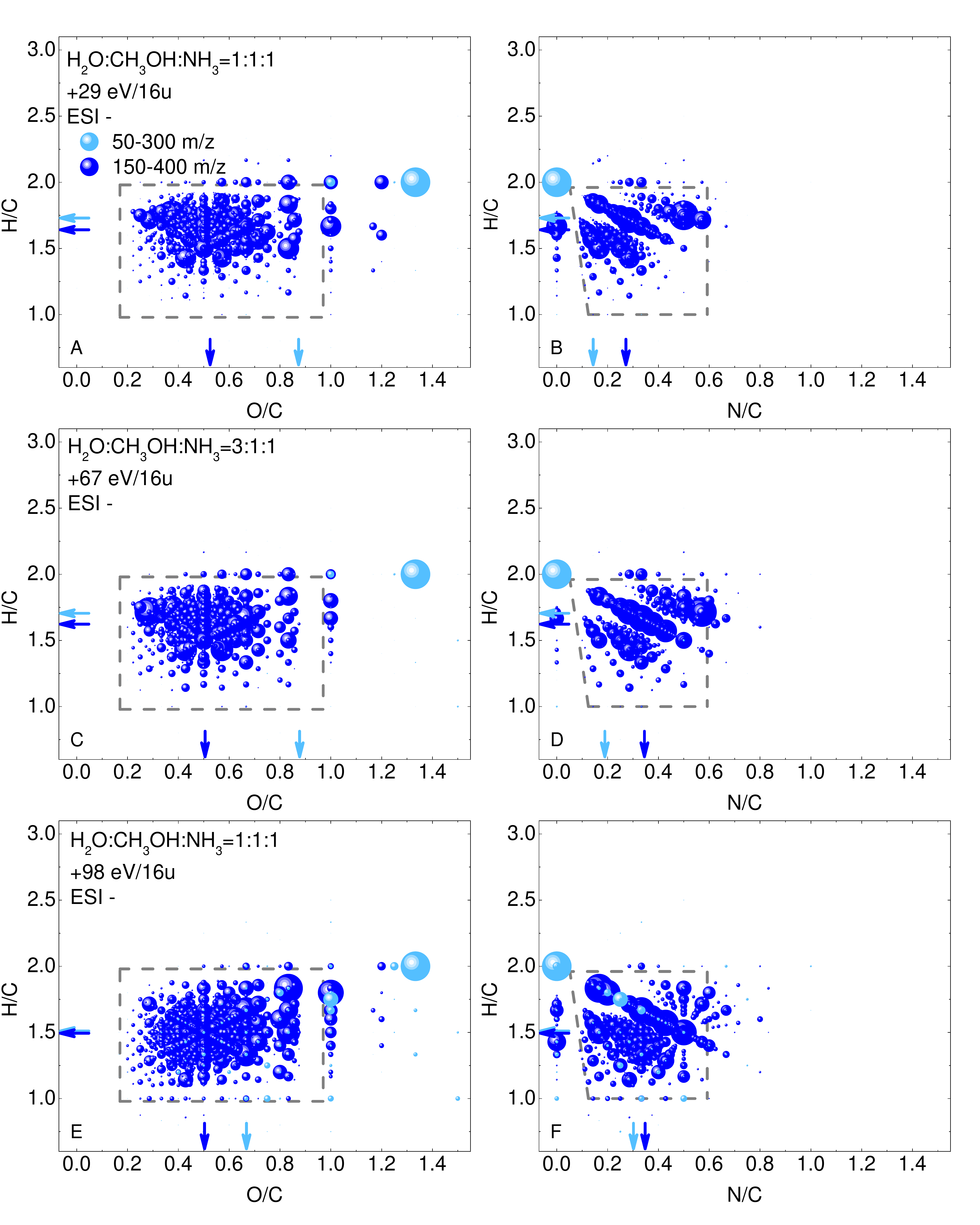}
	\caption{van Krevelen scatter plots of hydrogen to carbon versus oxygen to carbon (left panels) and versus nitrogen to carbon (right panels) in the three residues analysed in negative ESI mode in the mass ranges 50-300 (light-blue dots) and 150-400 (blue dots). Panels A and B: 1:1:1+29 eV/16~u residue; panels C and D: 3:1:1+67 eV/16~u residue; panels E and F: 1:1:1+98 eV/16~u residue. The size of the dots is given by the normalized intensity. Gray-dashed lines delimit the zone of highest density of ions observed in the residues analysed by \citet{danger16}. Arrows show the H/C, O/C and N/C weighted averages reported in Table~\ref{elemental_abu} for the range 50-300 m/z (cyan arrows) and 150-400 m/z (blue arrows).}
	\label{fig:figurehcocn}
\end{figure*}

All residues show the presence of three groups of molecules, namely CHO, CHN and CHNO. 
We calculate the weighted abundances of each group by dividing the sum of attribution intensities of a group by the total intensity of the dataset \citep{danger16}, assuming that the ionization yield is constant for each ion. 
For each residue, Fig.~\ref{fig.Figure_moleculargroups} shows the intensity of each group in the two ESI modes and in the two m/z ranges we investigate, i.e. 50-300 and 150-400 m/z, ranges where VHRMS allows an exact attribution of stoichiometric formulas. CHNOs are revealed in both positive and negative ESI mode, while CHOs and CHNs are detected in negative and positive ESI modes, respectively.
In the 1:1:1+98 eV/16~u residue, the CHNO group reaches the 94\% in abundance, followed by CHO (5.5\%) and CHN (0.3\%).
CHNOs dominate also in the 1:1:1+29 eV/16~u residue, where they account for the 79\% of the intensity of the whole dataset. In this sample, the CHO group has an abundance of about 17\%. CHOs are abundant in the 50-300 m/z range, while CHNs account for the 4\% of the dataset intensity. 
In the 3:1:1+67 eV/16~u residue, CHNOs account for the 82\% in abundance, followed by CHOs (about 15\%) and CHNs (slightly less than 4\%). 

In Fig.~\ref{fig:figurehcocp} and Fig.~\ref{fig:figurehcocn} we show van Krevelen plots of H/C versus O/C and N/C ratios. Plots include all the data in the ranges 50-300 and 150-400 m/z. Data at abscissa zero in the H/C versus O/C and H/C versus N/C diagrams belong to molecules that do not contain oxygen or nitrogen atoms, thus CHN and CHO, respectively. 
In the figure, the size of the dots is given by the intensity of each peak that is normalized with respect to the most intense peak in each m/z spectrum. 
For comparison, grey polygons show the areas with the highest density of ions observed in residues produced after UV photolysis and warm-up of H$_2$O:CH$_3$OH:NH$_3$ mixtures reported by \citet{danger16}, that use VHRSM in both positive and negative ESI modes to analyse such samples.
Plots reveal a good match between our samples and UV residues.
In Fig.~\ref{fig:OMass} and \ref{fig:NMass} we show the number of oxygen and nitrogen atoms for each formula with respect to the m/z ratio. 
The distribution of data in Van Krevelen plots suggest that samples show a homogeneous composition, and no difference attributable to the irradiation dose are displayed.

\begin{figure*}
	\centering
	\includegraphics[width=1\linewidth]{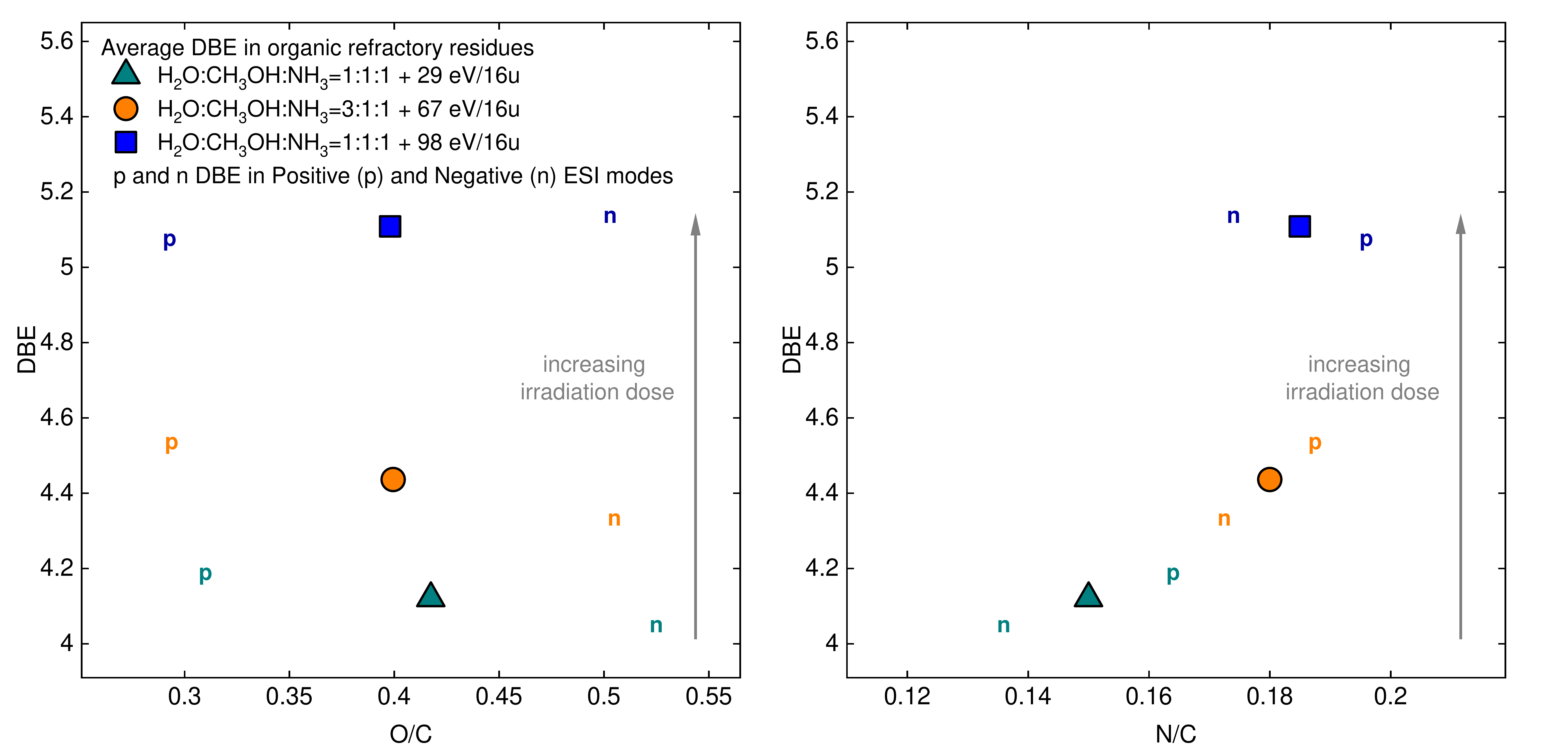}
	\caption{Double Bond Equivalent (DBE) versus O/C (left panel) and N/C (right panel) measured in the organic refractory residues produced after ion bombardment and warm-up of H$_2$O:CH$_3$OH:NH$_3$ mixtures in the range 150-400 m/z. "p" and "n" indicate the DBE values estimated in positive and negative mode, respectively, while squares, circles and triangles give the average values. Gray arrows indicate the evolution of DBE with increasing irradiation dose. We estimate Relative Standard Deviations of 2\% for DBE, 12\% for O/C and 6\% for N/C.}
	\label{fig.Figure_DBE}
\end{figure*}

In Fig.~\ref{fig.Figure_DBE} we show the weighted DBE with respect to O/C and N/C ratios calculated in both positive (p) and negative (n) ESI modes and in the range 150-400 m/z, a similar range to that used by \citet{fresneau17} to measure the DBE of UV residues produced after irradiation of H$_2$O:CH$_3$OH:NH$_3$ mixtures analysed through VHRMS in both ESI modes, i.e. 200-400 m/z.
In the figure, we also show the average of positive and negative ESI modes. Averaged DBE vary from 4.1 for the 1:1:1+29 eV/16 u to 4.4 for the 3:1:1+67 eV/16 u and 5.1 for the 1:1:1+98 eV/16 u. 
Irradiation dose seems to affect DBE values. In particular, DBE increases with increasing dose. 
We also calculate the unweighted DBE, thus following the same method used by \citet{fresneau17} in the estimation of the DBE of UV residues. Unweighted DBE also increase with increasing the irradiation dose. In fact, we obtain values of 4.54 for the 1:1:1+29 eV/16 u, 4.81 for the 3:1:1+67 eV/16 u and 5.52 for the 1:1:1+98 eV/16 u. Such values are consistent with those estimated by \citep{fresneau17} in UV residues.
For both DBE and elemental ratios, we estimate relative standard deviation (RSD) taking into account the values reported by \citet{fresneau17}, i.e. 6\% for N/C, 12\% for O/C, 3\% for H/C, and 2\% for DBE.
We note that we do not reveal differences induced by the higher amount of water in the 3:1:1 + 67 eV/16~u samples in van Krevelen plots, elemental abundances, and DBE.

\begin{figure*}
	\centering
	\includegraphics[width=0.9\linewidth]{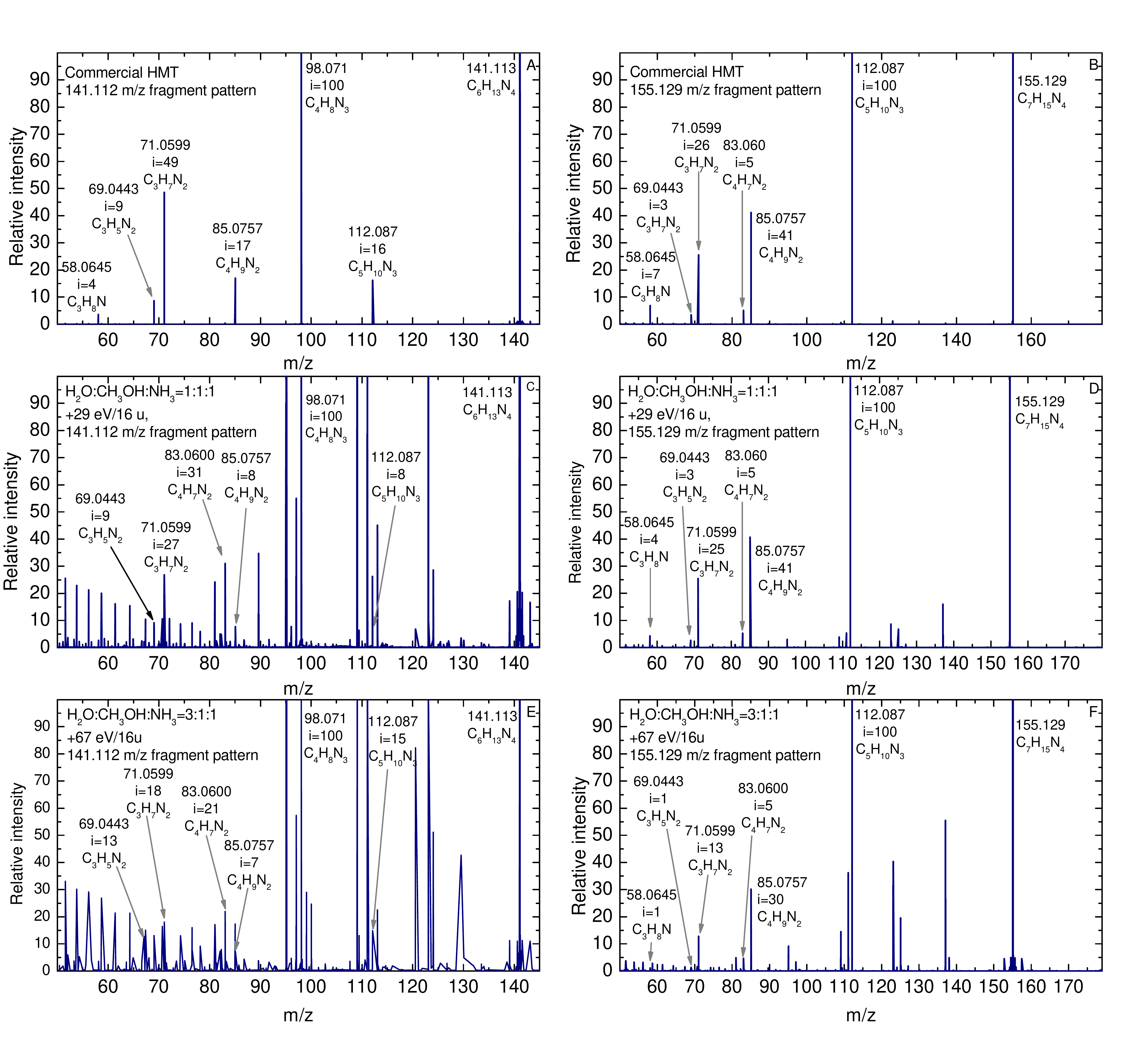}
	\caption{Fragment pattern of peaks at 141.113 (left panels) and 155.129 m/z (right panels) of: Panels A and B: commercial HMT; panels C and D: 1:1:1+29 eV/16~u residue; panels E and F: 3:1:1+67 eV/16~u residue. For each fragment, labels indicate the m/z value, the intensity with respect to the most intense fragment and the stoichiometric formula of the detected molecular ion M+H$^+$.}
	\label{fig:figurehmtfrag}
\end{figure*}

Further inspections of the mass spectra acquired in ESI positive mode reveal the presence of intense peaks at 155.129 m/z in all samples. We also detect peaks at 141.113 m/z. These peaks are attributed to compounds with the same stoichiometric formula of protonated HMT-CH$_3$ \citep[][]{danger13} and of protonated HMT. Our residues also display peaks at 185.140 m/z, also detected in UV residues and attributed to HMT-C$_2$H$_5$O or its isomers \citep{munozcaro04, danger13}.
The presence of these m/z peaks triggered us to investigate on the presence of HMT and its derivatives in our samples, especially because HMT has been revealed only once in residues produced after ion irradiation thanks to dedicated analysis reported by \citet{cottin01} and no information is given on how the HMT abundance varies with irradiation dose.
We thus performed further analysis by means of tandem MS/HRMS in order to obtain the fragment pattern of both m/z peaks by collision induced dissociation (CID) in the 1:1:1 +29 eV/16~u and in the 3:1:1+67 eV/16~u residues, and we compare the fragment pattern of the 141.113 peaks identified in residues with the fragment pattern of the 141.113 peak of pure commercial HMT. 
Our HMT sample also displays a peak at 155.129 m/z, thus we also have a fragment pattern of reference for the 155.129 m/z peaks. Furthermore, we compare our fragment pattern with those reported by \citet{danger13} that used VHRMS and MS/HRMS for the same purpose in UV residues (see Table~2 in their paper). The results of the analysis are shown in Fig.~\ref{fig:figurehmtfrag}. 
The intensities of the fragments of the 141.113 m/z molecular ions are normalized to the intensities of the 98.071 m/z peaks, while the intensities of the fragments of the 155.129 m/z molecular ions are normalized to the intensities of the 112.087 m/z peaks \citep{danger13}. 
All the molecular fragments identified in the MS/HRMS analysis of commercial HMT are revealed in both residues. Furthermore, the relative intensity of each m/z peaks we identify in residues is similar to the relative intensity of the same fragment identified in the commercial HMT and in the residues analyzed by \citet{danger13}.
We note that in the m/z spectra acquired in ESI positive mode, the intensity of m/z peaks associated to HMT and its derivatives decreases with irradiation dose as shown in Fig.~\ref{fig:figurehmtabundance}.

Other peaks in the m/z spectra shown in Fig.~\ref{fig:figurehmtfrag} are due to the presence of molecular ions with m/z 141 and 155 selected in the linear ion trap (resolution of 1 m/z) before the CID and the transfer of the fragments to the FT-orbitrap.
	In the two m/z ranges, we find peaks at 141.102 and 155.081 m/z whose intensity is higher than that of the 141.113 and 155.129 m/z peaks, respectively. The 141.102 (attributed to C$_7$H$_{13}$N$_2$O$^+$) and 155.081 (attributed to C$_7$H$_{11}$N$_2$O$_2^+$) m/z peaks are present also in the mass spectra of the residues shown in Fig.~\ref{fig.Figure_Orbitrap}, even if with a lower intensity with respect to the 141.113 and 155.129 m/z peaks.

\begin{table*}
	\caption{Intensity of m/z peaks of isomers of amino acids and nucleobases identified in residues in the range 50-300 m/z in ESI negative mode. Intensities are normalized to the most intense peak in m/z spectra acquired in the range 50-300 m/z in ESI negative mode, i.e. 105.019 m/z assigned to C$_3$H$_5$O$_4^-$. Isomers of (from top to bottom): histidine, glutamine, threonine, proline, aminoisobutanoic acid, alanine, glycine, thymine, cytosine, uracil.}	\label{prebioticintensity}
	\centering
	\begin{tabular}{cccc}
		\hline \hline
		Stoichiometric formula & \multicolumn{3}{c}{m/z peaks intensity}\\
		& 1:1:1 + 29 eV/16 u &3:1:1 + 67eV/16 u &1:1:1 + 98 eV/16~u\\ 
		\hline			
		C$_6$H$_9$N$_3$O$_2$& 0.24 &0.43	&5.8	\\		
		C$_5$H$_{10}$N$_2$O$_3$& 0.42	&0.60	&3.73\\	
		C$_4$H$_9$NO$_3$& 0.41&0.52		&3.27	\\					
		C$_5$H$_9$NO$_2$&0.44	&0.28&2.03\\					
		C$_4$H$_9$NO$_2$&0.12	&0.13	&1.51	\\	
		C$_3$H$_7$NO$_2$&0.22	&0.27	&3.16\\						
		C$_2$H$_5$NO$_2$&0.12	&0.38 	&2.89	\\			
		C$_5$H$_6$N$_2$O$_2$& 1.07	&1.30	&  14.63\\
		C$_4$H$_5$N$_3$O& 0.11	&0.35	& 3.31\\
		C$_4$H$_4$N$_2$O$_2$&	0.23&0.57	&5.91\\
		\hline
	\end{tabular}
\end{table*}

The analysis of m/z spectra reveals that other prebiotic compounds could be present in our samples. \citet{danger13} give a list of tentatively detected amino acids in residues produced after UV photolysis of H$_2$O:$^{13}$CH$_3$OH:NH$_3$ mixtures.
We search for the same stoichiometric formulas in our residues, and we identify the m/z peaks of molecular ions attributed to the following amino acids or their isomers: histidine (C$_6$H$_9$N$_3$O$_2$,  ESI - 154.062 m/z, ESI + 156.077 m/z), glutamine (C$_5$H$_{10}$N$_2$O$_3$, ESI - 145.062 m/z, ESI+ 147.076 m/z), threonine (C$_4$H$_9$NO$_3$, ESI- 118.051 m/z, ESI+ 120.065m/z), proline (C$_5$H$_9$NO$_2$, ESI- 114.056 m/z), amino isobutanoic acid (C$_4$H$_9$NO$_2$, ESI- 102.056 m/z, ESI+ 104.071 m/z), alanine (C$_3$H$_7$NO$_2$, ESI- 88.0404 m/z), and glycine (C$_2$H$_5$NO$_2$, ESI- 74.0247 m/z). 
We also identify m/z peaks attributed to pyrimidine nucleobases or their isomers, i.e. uracil (C$_4$H$_4$N$_2$O$_2$, ESI - 111.020 m/z), cytosine (C$_4$H$_5$N$_3$O, ESI- 110.036, ESI+112.051 m/z) and thymine (C$_5$H$_6$N$_2$O$_2$, ESI- 125.036 ). None of the residues here analysed show stoichiometric formulas of purine nucleobases, adenine and guanine. 
Table~\ref{prebioticintensity} reports the relative intensity with respect to the most intense m/z peak in each spectrum (i.e. C$_3$H$_5$O$_4^-$, 105.019 m/z) of tentatively detected amino acids and nucleobases stoichiometric formulas.
We note that the relative intensity of these m/z peaks increases with the irradiation dose given to the frozen mixture.
However, despite the procedure we performed to get rid of contamination described in Sect.~\ref{experimentalmethods}, we cannot exclude the presence of contaminants that contribute to the detected signals. Further analysis are needed to investigate the presence and the dependence on the irradiation dose of prebiotic compounds using both isotopic labelling and MS/HRMS analysis of the fragmentation pattern.

\section{Discussion}
\subsection{The composition of organic refractory residues and the effects of dose}
The analysis here reported gives information on the chemical composition of organic refractory materials produced after ion irradiation of frozen volatiles. 
By means of IR spectroscopy, we identify various bands attributed to several functional groups, as reported in Table~\ref{residues}. 
The high molecular diversity of such samples is better highlighted through VHRMS. This technique reveals the presence of thousands of molecular ions in each sample and three diverse group of molecules, namely CHO, CHN and CHNO, with the latter being the most abundant (up to 94\% of all attributions in the 50-400 m/z range).

In Table~\ref{elemental_abu} we calculate the elemental abundances as H/C, N/C and O/C, and we observe variations that we relate to the irradiation dose given to the pristine frozen mixture. In Fig.~\ref{fig:figureelemabuvsdose}, H/C and N/C ratios are also plotted with respect to the irradiation dose given to the frozen mixtures. 
We observe that the H/C ratio decreases while the N/C ratio increases with increasing the dose. 
Taking into account the estimation reported by \citet{strazzulla03}, we also show the timescales needed to accumulate doses that affect elemental abundances at different heliocentric distances through irradiation by CR and SEP within the upper 50 $\mu$m of a frozen surface. Such timescales are given for SEP and CR at 40 AU and for CR at the termination shock (85 AU) and in the Local Interstellar Medium (LISM).

The decrease of the H/C with respect to the dose has been reported in the case of the ion irradiation of benzene by \citet{strazzullabaratta92}, due to the loss of H$_2$ and a progressive carbonization during irradiation. The loss of hydrogen has also been observed in hydrogenated amorphous carbon exposed to irradiation \citep{mate14}.
N and C elemental abundances of residues produced after ion irradiation have been reported by \citet{accolla18}.  
Even though they used different initial mixtures (with initial N/C ratios between 0.25 and 2 with respect to our initial N/C=1), taking into account the data reported in Table~2 in \citet{accolla18}, they observe an increase of the N/C ratio with increasing dose. 
Together with H/C and N/C ratios, also DBE show variations related to the dose given to the frozen mixture, and DBE increases with increasing dose. 
Elemental abundances, and consequently DBE, could be also affected by changes in the chemical composition of residues induced by their temporal evolution or the exposure to atmosphere, during the removal of residues from the vacuum chamber and during their storage. 
\citet{accolla18} and \citet{baratta19} monitored the evolution of residues through IR spectroscopy. Their stabilization can require up to 200 days and it is induced by sublimation of volatiles and reactions within the residue matrix.
Our samples are stored under vacuum, and there is no evidence of alterations induced by the exposure to atmosphere, as it would determine an increase in the O/C ratio of residues.
In particular, we would expect higher O/C ratios in the 1:1:1+98 eV/16 u residue, produced 240 days before analysis, than in the other samples, analyzed only 2 weeks after their production. However, the 1:1:1+98 eV/16~u residue shows similar O/C ratios (and even lower, in the range 50-300 m/z) than the other two samples.

A detailed comparison between our samples and UV residues it outside the main scopes of this work. However, VHRMS in both positive and negative ESI modes has been used to analyse UV residues, and we now gather the information obtained on both types of samples. 
For UV residues, \citet{danger16} show elemental abundances and van Krevelen plots, and \citet{fresneau17} estimate their unweighted DBE. The distribution of data in van Krevelen plots that we show in Fig.~\ref{fig:figurehcocp} and Fig.~\ref{fig:figurehcocn} resemble the distribution observed for UV residues, implying a similarity in the chemical composition of the residues, despite the different source of processing. Also unweighted DBE are found to be similar in both type of residues.
For what concerns elemental abundances, taking into account the results of this work, the higher H/C and lower N/C observed in UV residues can be explained by the lower UV irradiation dose given to the pristine ice, i.e. about 10~eV/16 u. 

A common characteristic of UV residues is the presence HMT and its derivatives, whose main vibrational mode bands are found between 1250-950~cm$^{-1}$.
In the IR spectra shown in Fig.~\ref{fig:figureres1300-900} we do not find clear evidence of the presence of such compounds, because any attribution is hindered by the presence of intense features in a N-free residue (1:1:0 +85 eV/16 u), where no HMT can be present.
Tandem MS/HRMS allowed the detection of stoichiometric formulas tentatively attributed to HMT and its derivatives in residues. 
The fragment pattern of peaks at 141.113 and 155.129 m/z are similar to those revealed in commercial HMT and to the fragment pattern reported by \citet{danger13}, attributed to HMT and HMT-CH$_3$, respectively. 
The m/z spectra of residues also show peaks at 185.140 m/z, attributed to HMT-C$_2$H$_5$O or its isomers and previously revealed in UV residues \citep{danger13}.
Irradiation dose plays a role not only in the changes observed in elemental abundances and DBE values, but also in the production of HMT.
In fact, the intensity of m/z peaks tentatively attributed to HMT and its derivatives diminish as the irradiation dose given to the frozen mixtures increases, as shown in Fig.~\ref{fig:figurehmtabundance}. 
An explanation could be found in the destruction of the HMT precursors produced by irradiation in the frozen matrix, i.e. H$_2$CO and HCOOH. High irradiation doses would partially destroy the newly formed species, and as a result, lower quantities of HMT could be formed throughout the warm up.
This consideration is supported by the fact that the 1:1:1+29 eV/16 u residue, the one produced at the lowest irradiation dose, displays the most intense m/z peaks associated to stoichiometric formulas of HMT and its derivatives, and its IR spectrum shows a shoulder at about 1010~cm$^{-1}$ that could be attributed to HMT. 

Finally, in m/z spectra of residues we note the presence of stoichiometric formulas attributed to isomers of amino acids and nucleobases, and their intensity seems to be affected by the extent of irradiation dose. In future, dedicated analysis should be performed to deepen the effects of dose in the production of prebiotic species.

\subsection{Astrophysical and Astrobiological implications}
Water, methanol and ammonia are widely present in the ISM as well as in the solar system. 
In the outer solar system, water ice has been detected on the surface of various small bodies \citep[e.g.][]{barucci11}. Observations and space mission data also revealed the presence of methanol on the surface of 5145~Pholus \citep{cruikshank98}, 55638~2002 VE$_{95}$ \citep{barucci06} and (4869580)~Arrokoth \citep{grundy20}. 
So far, NH$_3$ has been revealed through its 2.22~$\mu$m band in localized areas on the surface of Charon \citep{grundy16}, and \citet{delsanti10} suggested that the irradiation of NH$_3$ is responsible for the formation of NH$_4^+$ on the KBO (90482)~Orcus, and it could be present in the interior of outer bodies \citep{lisse20}. 
Recently, \citet{altwegg20} and \citet{poch20} suggested the presence of ammonium salts on the comet 67P/Churyumov-Gerasimenko. According to \citet{poch20}, these salts would contribute to the 3.2~$\mu$m feature detected on the comet surface, together with water ice and organics, and ammonium salts would originate thanks to reactions involving ammonia.
Furthermore, experiments reported by \citet{parent09} revealed that irradiation of ammonia produces up to 12\% of N$_2$, the latter being revealed on Pluto \citep{owen93, grundy16}, 90377~Sedna \citep{emery07}, 136199~Eris \citep{tegler12}, and on the Neptune satellite Triton \citep{cruikshank93}.
Thus, the analysis of organic residues produced after the irradiation of water, methanol and ammonia mixtures is necessary to shed light on the physical and chemical properties of cometary nuclei and of the surfaces of small bodies in the outer solar system, so far being accessible only by means of observations and data acquired by the New Horizons space probe.

In this work, we link the composition of organic refractory materials to the irradiation dose given to the pristine frozen mixture.
Recently, \citet{urso20} found that irradiation doses capable to induce detectable changes in the spectra of frozen surfaces, i.e. about 40~eV/16~u, would be accumulated within 10$^6$-10$^7$ years, according to the heliocentric distance. Such short timescale testify the efficiency of irradiation in modifying the physico-chemical properties of frozen surfaces in the outer solar system. In particular, \citet{urso20} show the primary role of SEP in affecting frozen surfaces in the outer solar system. 
In fact, the timescales reported in Fig.~\ref{fig:figureelemabuvsdose} show that in the Kuiper-belt, at the current orbit of Arrokoth, between 2.1 and 7.4$\times$10$^7$ years would be needed to accumulate doses that affect elemental abundances thanks to the irradiation by SEP. 
At the same location, the heliosphere shield CR and to accumulate the same dose CR irradiation would require up to 8$\times$10$^9$ years. However, CR become the main source of processing beyond the termination shock ($\sim$85 AU), and in the Local Interstellar Medium the timescale of irradiation by CR shorten to 10$^6$-10$^7$ years.
We note that the timescale of irradiation by solar particles would further shorten taking into account that the young sun might have been more active than it is at present times \citep[e.g.][]{gudel97}.

The formation of organic refractory materials on frozen surfaces requires both irradiation by energetic particles and heating. After being exposed to energetic particles in the outer solar system, small bodies can migrate inward the solar system and experience heating, allowing the formation of organic refractory materials. 
In laboratory, we observe the formation of organic refractory residues at about 300~K.
However, in space both diffusion and sublimation of volatiles needed to form organic refractory materials can happen in timescales of billions of years. Thus, such refractory materials might form even at lower temperatures.

Kuiper-belt objects are thought to have formed in the protoplanetary nebula \citep[e.g.][]{grundy20}. 
During their formation, small bodies could have accreted organic molecules present in the presolar disk, possibly produced thanks to both solid and gas-phase reactions in the interstellar medium. However, the harsh conditions of the protoplanetary disk could have determined the destruction of organics and complex compounds. 
Thus, a possibility is that only simple species were available to be accreted in the seeds of planetesimals. 
In this scenario, our data suggest that the production of organic molecules through ion bombardment would have required only a few tens of million years after the formation of a frozen body. The high extent of molecular diversity revealed in our sample may thus exist on frozen surfaces in the outer solar system. 

Our analysis also point out that ion irradiation favours the formation of compounds of interest for astrobiology.
After ion bombardment, at 15~K we observe various bands attributed to formamide, a compound revealed in star-forming regions \citep[e.g.][]{lopezsepulcre15} as well as in comets \citep{bockelee00, biver14, goesmann15} that has a relevant role in the synthesis of various prebiotic compounds, including nucleobases \citep[e.g.][]{saladino12, ferus14, kanuchova16, urso17, botta18, lopezsepulcre19}.
In residues, we detect m/z peaks tentatively attributed to HMT and HMT derivatives. HMT is a precursor of various organic compounds, including amino acids \citep[e.g.][]{hulett71, bernstein95, vinogradoff20} and its formation after irradiation and heating of astrophysical ice analogues suggests that it could be present in solar system outer objects. Furthermore, in residues we reveal the presence of amino acids and nucleobases or their isomers. 
In future, the instruments on board the James Webb Space Telescope will allow to observe the wavelength range where HMT shows its most intense bands. In particular, the Mid-infrared Instrument (MIRI) will cover the range 5-28~$\mu$m (2000-360~cm$^{-1}$). 
As reported in this work, HMT shows intense vibrational mode bands in the same wavelength range where also N-free residues exhibit intense bands (see Fig.~\ref{fig:figureres1300-900}). As a result, caution must be taken when attributing bands to HMT in this wavelength range.

\section{Conclusions}

The frozen surfaces of outer objects in the solar system show the presence of various volatile molecules as well as refractory red materials, whose presence may be due to the effect of the bombardment by cosmic rays, solar energetic particles and solar wind.
To date, only limited information is available on the composition of organic refractory materials on the surfaces of frozen bodies in the outer solar system, as astronomical observations only show the presence of red slopes in the visible and near-infrared.

Our experiments simulate the formation of refractory materials on such frozen surfaces, and shed light on their composition. 
Analysis reveal the high extent of molecular diversity of residues. Very High Resolution Mass Spectrometry data show the presence of CHO, CHN, and CHNO molecular groups. The distribution of elemental ratios reported in van Krevelen plots suggests a compositional homogeneity of residues, with distributions similar to those revealed for UV residues. 
According to our data, elemental abundances, and in particular the H/C and N/C ratios are affected by the dose given to the deposited frozen volatile mixture. 
Increasing the dose determines a lowering of the H/C ratio, while the N/C ratio increases. The dose also affects the Double Bond Equivalent. In fact, the level of unsaturation of residues increases with the increasing dose. 

Although the presence of hexamethylenetetramine in our samples cannot be confirmed with IR spectroscopy, by means of tandem Mass Spectrometry/High Resolution Mass Spectrometry we reveal a fragment pattern that corresponds to that of hexamethylenetetramine, and for the first time, we report the presence of stoichiometric formulas associated to its derivatives in residues obtained after ion irradiation of frozen volatiles. Furthermore, data suggest a dependence of the abundance of hexamethylenetetramine and its derivatives to irradiation dose.
The detection of such compounds in our samples opens to the possibility that ion irradiation, and not only UV photolysis, induces their formation, in agreement with the results reported by \citet{cottin01}. Thus, hexamethylenetetramine could be present also in astrophysical environments shielded from UV radiation. 
Taking into account the effects induced by solar energetic particles on frozen surfaces in the outer solar system, we estimate that doses capable to produce organic refractory materials with a composition similar to that of the samples here analyzed would be accumulated within 2-7$\times$10$^7$ years on a frozen surface located at about 40 AU.

\begin{acknowledgements}
We thank the anonymous Referee and the Editor for their comments. 
We thank D. Ledu and C.O. Bacri for the access to the SIDONIE facility. INGMAR is a IAS-IJCLab facility funded by the French Programme National de Plan\'{e}tologie (PNP), Facult\'{e} des Sciences d'Orsay, Universit\'{e} Paris-Sud (Attractivit\'{e} 2012), P2IO LabEx (ANR-10-LABX-0038) in the framework Investissements d'Avenir (ANR-11-IDEX-0003-01). 
We thank the support from RAHIIA SSOM (ANR-16-CE29-0015). This work is supported by the French National Research Agency in the
framework of the Investissements d'Avenir program (ANR-15-IDEX-02), through the funding of the “Origin of Life” project of the Universit\'{e}
Grenoble Alpes and the French Space Agency (CNES) under their Exobiology and Solar System programs. We thank D. Baklouti for fruitful discussions.
R.G.U. thanks the CNES postdoctoral program. 
\end{acknowledgements}

\bibliography{riccardo_bib.bib}
\bibliographystyle{aa}

\begin{appendix}
	
	\section{Vibrational mode bands identified in processed ices and in organic refractory residues}
	In the following tables, we list the vibrational mode bands detected after irradiation with 40 keV H$^+$ of H$_2$O:CH$_3$OH:NH$_3$ mixtures at 15~K (Table~\ref{ice_irradiation}) and those in organic refractory residues at 300~K (Table~\ref{residues}). Vibrational mode bands are attributed taking into account the results of similar irradiation experiments reported in literature.
	
	\begin{table*}
		\caption{Bands at 15 K after irradiation with 40 keV H$^+$ of H$_2$O:CH$_3$OH:NH$_3$ mixtures. v int= very intense, int= intense, s=shoulder, w=weak, b=broad, n= noisy.}	\label{ice_irradiation}
		\centering
		\begin{tabular}{lllccc}
			\hline \hline
			1:1:1&3:1:1&1:1:1 &Mode  & Compound & Ref.\\
			+29 eV/16~u&+67 eV/16~u&+98 eV/16~u& &&\\
			\hline 
			3378 (v int)	&3373 (v int)	&3375 (v int)   	&$\nu$(NH) &NH$_3$							&1\\
			3261 (v int, b)&3261 (v int, b)&3272 (v int, b)	&$\nu$(OH) &H$_2$O, CH$_3$OH			&2\\
			3009 (w)			&	3013 (w)	&3012 (w) 		&$\nu$(CH)&CH$_4$								&3\\
			2984 (w)		&	2990 (w)	&2980 (w) 		& 		&C$_2$H$_6$								&4\\
			2958 (w)		& 2960 (w)		&- 				& 	&CH$_3$OH									&2\\
			2938 (w)		&2938 (w)		&2936 (vw) 		&	 	&CH$_3$OH									&2\\
			2908 (w)		&2909(w)		&2900 (vw)		&	 	&H$_2$CO									&5\\
			2830 (int)		&2832 (int)		&2830 (w)		&	 	&CH$_3$OH, C$_2$H$_6$					&3, 4\\
			2343 (v int)	&2343 (v int)	&2343 (v int)	&$\nu$(CO)	&CO$_2$ 					&6\\
			2277 (w)		&2278 (w)		&2278 (int)		&	$\nu$(CO)&	$^{13}$CO$_2$						&7\\
			2260 (vw)		&2261 (vw, s)	&2261 (b)		&$\nu$(N=C=O)&HNCO							&8\\
			2241 (vw)		&2242 (w)		&2242 (b)		&	$\nu_3$&N$_2$O								&8\\
			2165 (int)		&2168 (int)		&2168 (int)		&	$\nu$(CN)&OCN$^-$							&2\\
			2138 (int)		&2139 (int)		&2140 (int)		&$\nu$(C$\equiv$O)&	CO						&9\\
			1846 (w) 		& 1845(w)	 	&1845 (w)		&$\nu$(CO)& HCO								&1\\
			1718 (int)		&1717 (s)		&1718 (s)		& $\nu$(CO) &H$_2$CO, HCOOH, NH$_2$HCO, 			&\\
			&				&				&			&(CH$_3$)$_2$CO, HCOOCH$_3$			&10, 11, 12\\ 
			1694 (s)		&1690 (v int	&1690 (v int)	& $\nu$(CO)&NH$_2$HCO 				&13\\ 
			1640 (int)		& 1643 (int, s)	&1640 (int, s)	& $\delta$(OH), (NH)& H$_2$O						&10, 14\\ 
			1591 (int)		&1591 (int)		&1591 (int, s)	& $\nu_{asy}$(COO$^-$)&HCOO$^-$			&10\\
			1500 (int)		&1500 (int)		&1499 (int)		& $\delta$(CH$_2$)    &H$_2$CO		&10\\
			1479 (int)		&1479 (s)		&1479 (s)		&				&NH$_3$, NH$_4^+$					&15\\ 
			1463 (int)		&1463 (s)		&1460 (s)		& $\delta$(OH), (NH)& CH$_3$OH, NH$_4^{+}$		&10\\
			1385 (int)		&1385 (int)		&1386 (int)		&$\delta_{sym}$(CH) &HCOO$^-$, NH$_2$HCO		&10, 8\\
			1352 (int)		& 1353 (int)	&1353 (int)		&   $\delta_{asy}$(CH)&HCOO$^-$, CH$_3$CHO	&10, 4, 15\\
			1305 (int)		& 1305 (int)	&1305 (int)		&   			& CH$_4$  						&4\\
			1250 (w)		&1247 (vw)		&1250 (vw)		& $\rho$(CH$_2$)   &H$_2$CO						&10\\
			1221 (vw)		& 1222 (vw)	&1218 (w, s)		& 	& (CH$_3$)$_2$CO								&11\\
			1125 (int)		&1126 (int)		&1124 (int)		&$\omega$(NH), $\rho$(CH)&NH$_3$, CH$_3$OH	&10, 6\\
			1095 (int, s)	&1094 (w, s)	&1094 (w, s)  		&				& (CH$_3$)$_2$CO					&11\\
			1029 (int)		&1024 (int)		&1025 (w)		& $\nu$(CO) 	& CH$_3$OH						&10\\
			820 (b)			& 820 (b) 		& 820 (b)		& libration			&H$_2$O						&10\\
			\hline	    	
		\end{tabular}
		\tablebib{1 \citet{d'hendecourt86}, 2 \citet{islam14}, 3 \citet{ferini04}, 4 \citet{moorehudson98}, 5 \citet{bossa09}, 6 \citet{palumbostrazzulla93}, 7 \citet{strazzulla99}, 8 \citet{kanuchova16}, 9 \citet{urso16}, 10 \citet{vinogradoff13}, 11 \citet{baratta94}, 12 \citet{modicapalumbo10}, 13 \citet{urso17}, 14 \citet{vinogradoff11}, 15 \citet{raunier04}.}
	\end{table*}

	\begin{table*}
		\caption{Vibrational bands observed in the spectra of organic refractory residues after 30 minutes at 300~K. v int= very intense; int=intense, s=shoulder; w=weak; b=broad, n=noisy}\label{residues}
		\centering
		\begin{tabular}{ccccc}
			\hline \hline
			\multicolumn{5}{c}{Organic refractory residues at 300~K}\\
			1:1:1&3:1:1&1:1:1&&\\
			+29 eV/16~u&+67 eV/16~u&+98 eV/16~u &Mode&Ref.\\
			\hline
			3306 (int, b)	&3197 (int, b) &3277 (int, b)&$\nu$(OH), $\nu$(NH)	&1\\
			2975 (s)			&2970 (s, n)&2976 (int)		&$\nu$(CH)				&1\\
			2936 (int)			&2936(s)	&2940 (int)		& $\nu$(CH)$_{sym}$		&2\\
			2878 (int)			&2872 (s, b)&2879 (int)		&$\nu$(CH)$_{asy}$		&3,2\\
			2216(w)			&2217(w,b)	&2215 (int)		&$\nu$(C$\equiv$N)			&1\\
			2161 (vw, b)		&2160(w, b)	&2161 (int)		&$\nu$(N$\equiv$C)		&1\\
			1749 (int)			&1739 (int)	&1746 (s)  		&$\nu$(C=O)				&4\\
			1720 (s)			&1720 (s)	&1720 (s)		&	$\nu$(C=O)			&4, 5\\
			1674 (int)			&1678 (int)	&1674 (v int)	&$\nu$(C=O)				&3\\
			1643 (int, n)		&1650 (int)	&1650 (s)		&$\nu$(C=O), (C=N), $\delta$(NH)&2\\
			1599 (int)			&1590 (b)	&1593 (int)		&$\nu$(COO$^-$), (C=C), (C=N)	&3, 5, 6\\
			1456 (s)			&1457 (b) 	&1458 (b)		& $\delta$(CH, NH)	&3, 4\\
			-					&1376 (w)	& 1382 (w)		&$\delta$(CH)			&5	\\	
			1345 (vw)			&1340 (s, w)&1343 (w)		& $\nu_s$(COO$^-$)		&2\\
			1232 (vw)			& 1234 (int)&1230 (int)		&$\nu$(C-N)				&3\\
			1111 (s)			&1110 (s)	&1110 (s, b)		&	$\nu$(C--O)		&5\\
			-					&1074 (int)	&1078 (int)		& 						&4\\
			1056 (v int)		&1050 (s)	&1050 (s)		&$\nu$(C--O)			&5\\	
			1009(s, w)			&1014 (s, w)&-				&$\nu$(C--N)				&3\\
			925 (vw)		& 926 (vw)	&925	(vw)		&	O--H def.		&4\\
			826 (int, b) 		&825 (s, b)	&827	(w, s)		& $\omega$(NH)		&4\\
			\hline	
		\end{tabular}
		\tablebib{1 \citet{accolla18}, 2 \citet{fresneau17}, 3 \citet{vinogradoff13}, 4 \citet{munozcaro03}, 5 \citet{bernstein95}, 6 \citet{demarcellus17}, 7 \citet{briani13}.}
	\end{table*}

	\section{Oxygen and nitrogen elemental abundances}
	
\begin{figure*}
		\centering
		\includegraphics[width=0.99\linewidth]{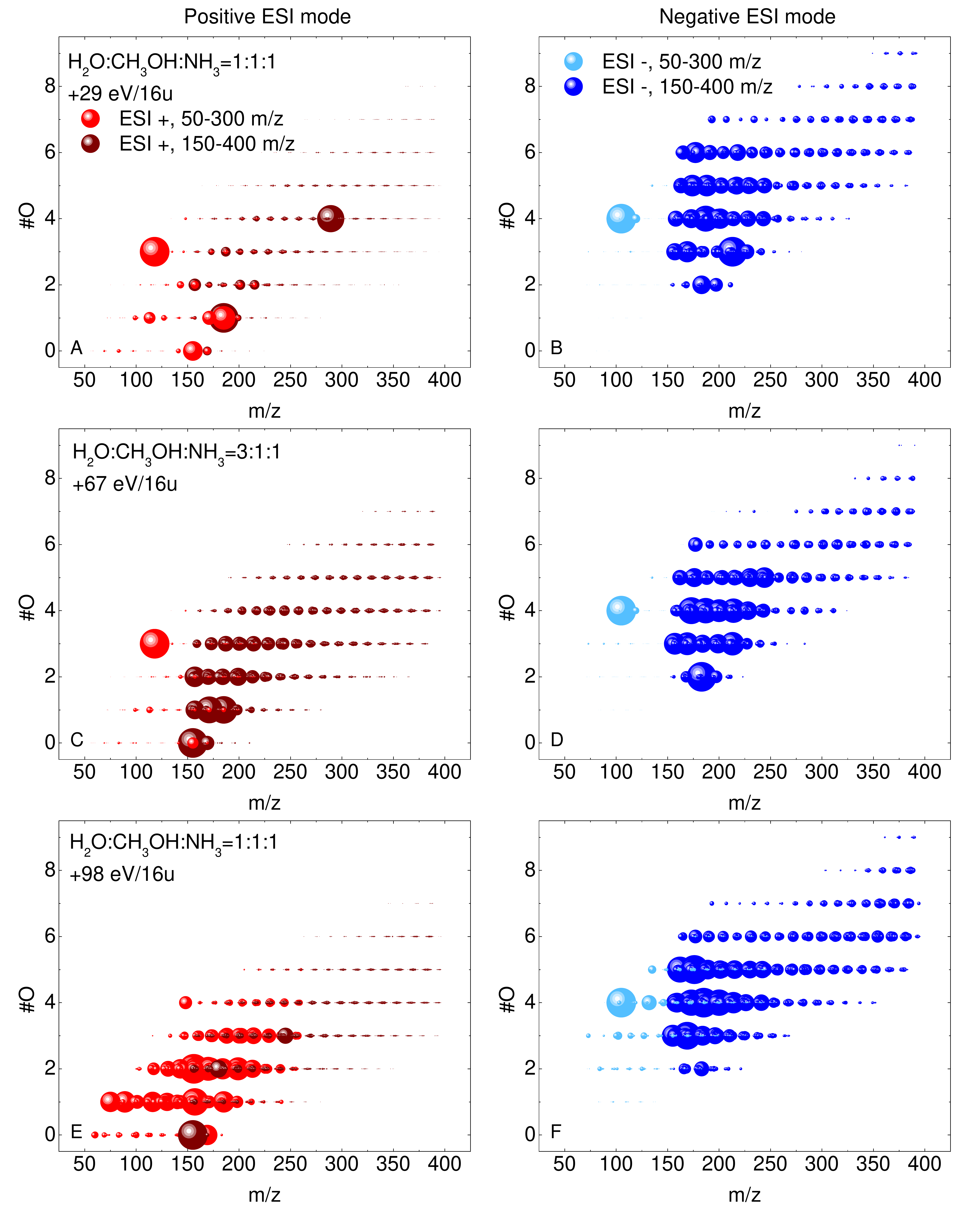}
		\caption{Number of oxygen atoms in stoichiometric formulas with respect to the m/z ratio in the range 50-300 and 150-400 m/z in both positive ESI mode (left panels) and negative ESI mode (right panels). Panels A and B: 1:1:1+29 eV/16~u residue; panels C and D: 3:1:1+67 eV/16~u residue; panels E and F: 1:1:1+98 eV/16~u residue. The size of the dots is given by the normalized intensity.}
		\label{fig:OMass}
	\end{figure*}

\begin{figure*}
	\centering
	\includegraphics[width=0.99\linewidth]{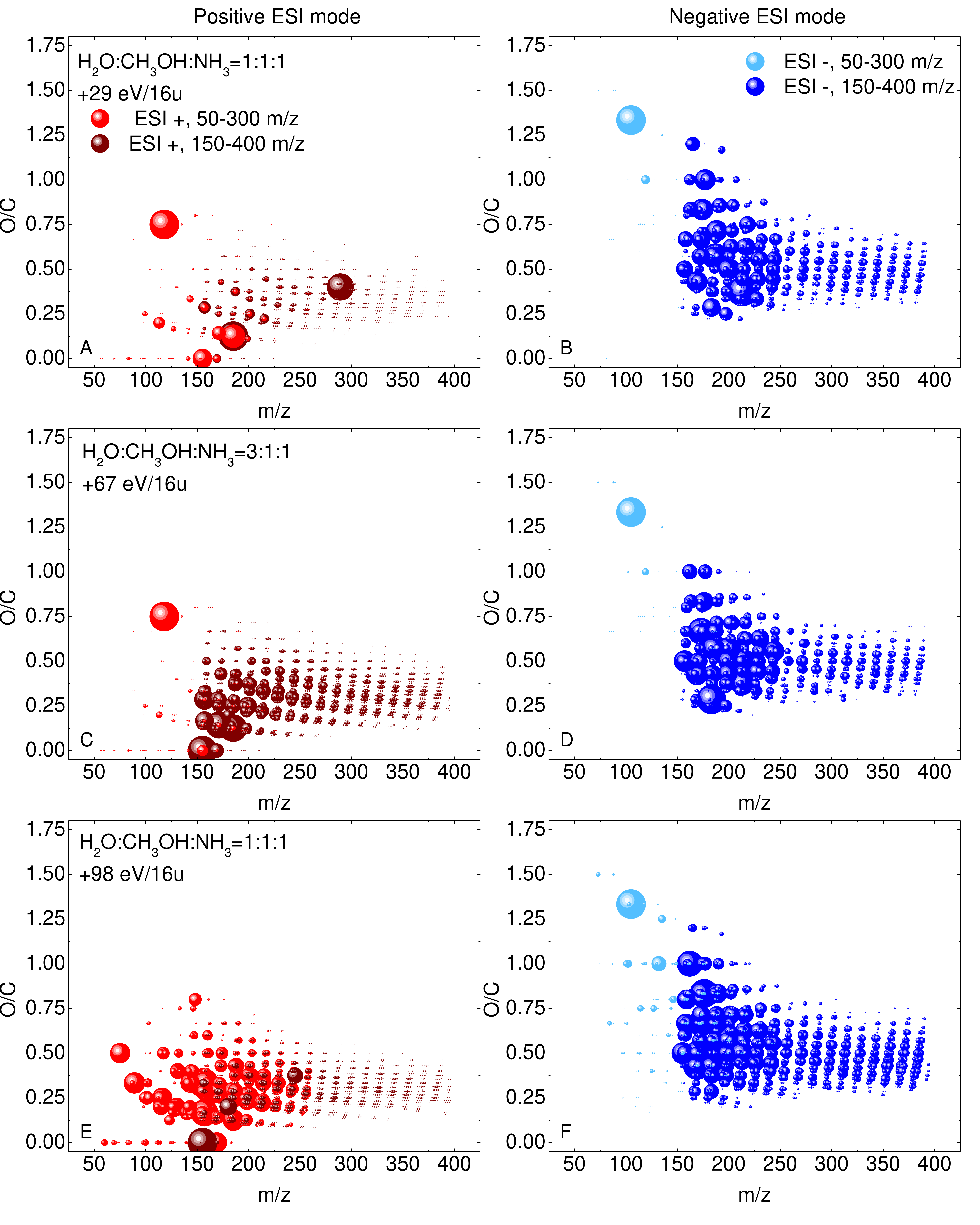}
	\caption{O/C ratios with respect to the m/z ratio in the range 50-300 and 150-400 m/z in positive ESI mode (left panels) and negative ESI mode (right panels). Panels A and B: 1:1:1+29 eV/16~u residue; panels C and D: 3:1:1+67 eV/16~u residue; panels E and F: residue of the 1:1:1+98 eV/16~u. The size of the dots is given by the normalized intensity.}
	\label{fig:OCMass}
\end{figure*}

	Figure~\ref{fig:OMass} and Fig.~\ref{fig:OCMass} show the elemental abundance of oxygen as the number of O atoms and the O/C ratio with respect to the m/z. The interpretation of the different distribution of intensities in the ranges 50-300 and 150-400 m/z is given in sect.~\ref{VHRMS}. 
	
\begin{figure*}
		\centering
		\includegraphics[width=0.99\linewidth]{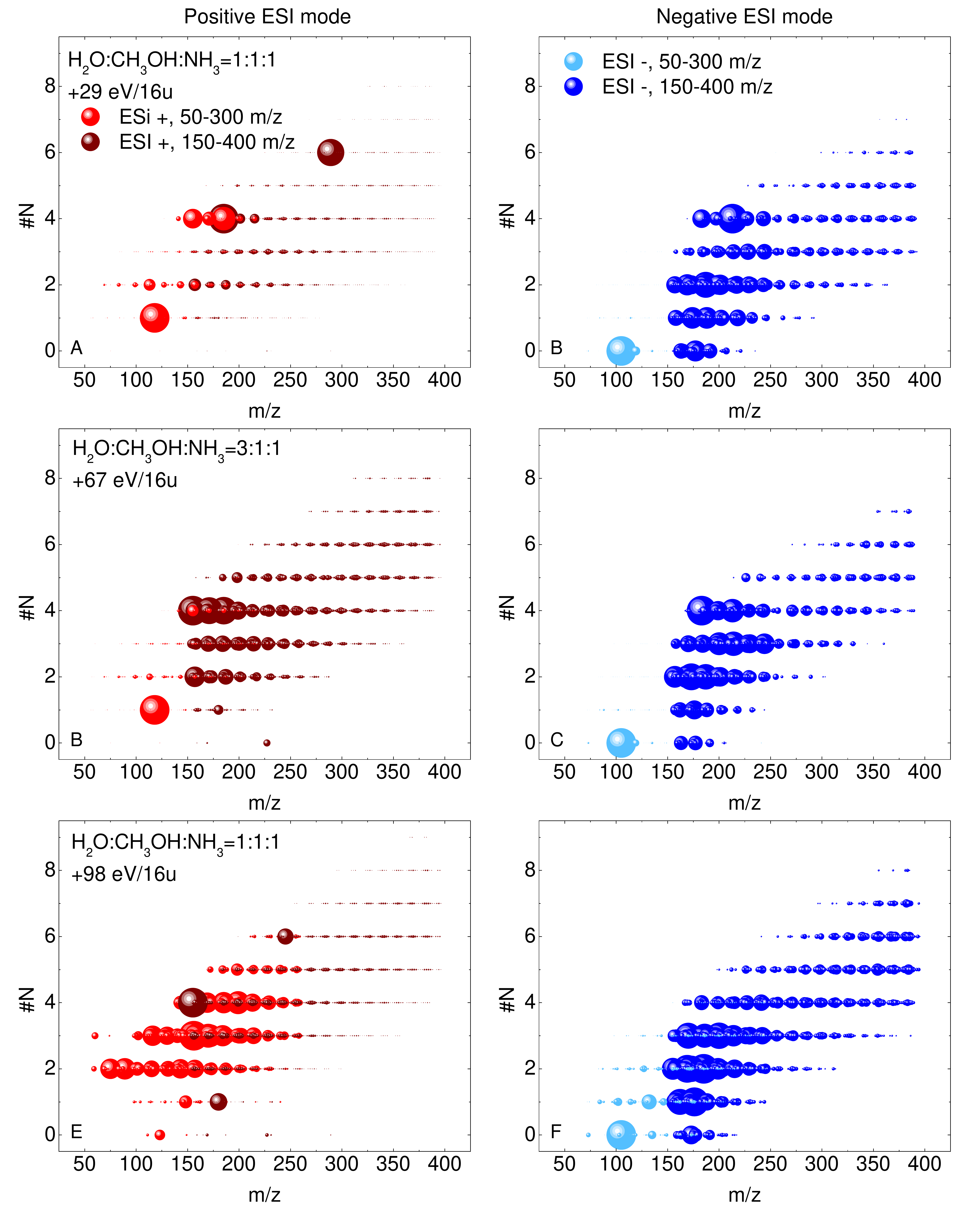}
		\caption{Number of nitrogen atoms in stoichiometric formulas with respect to the m/z ratio in the range 50-300 and 150-400 m/z in positive ESI mode (left panels) and negative ESI mode (right panels). Panels A and B: 1:1:1+29 eV/16~u residue; panels C and D: 3:1:1+67 eV/16~u residue; panels E and F: 1:1:1+98 eV/16~u residue. The size of the dots is given by the normalized intensity.}
		\label{fig:NMass}
	\end{figure*}

\begin{figure*}
		\centering
		\includegraphics[width=0.99\linewidth]{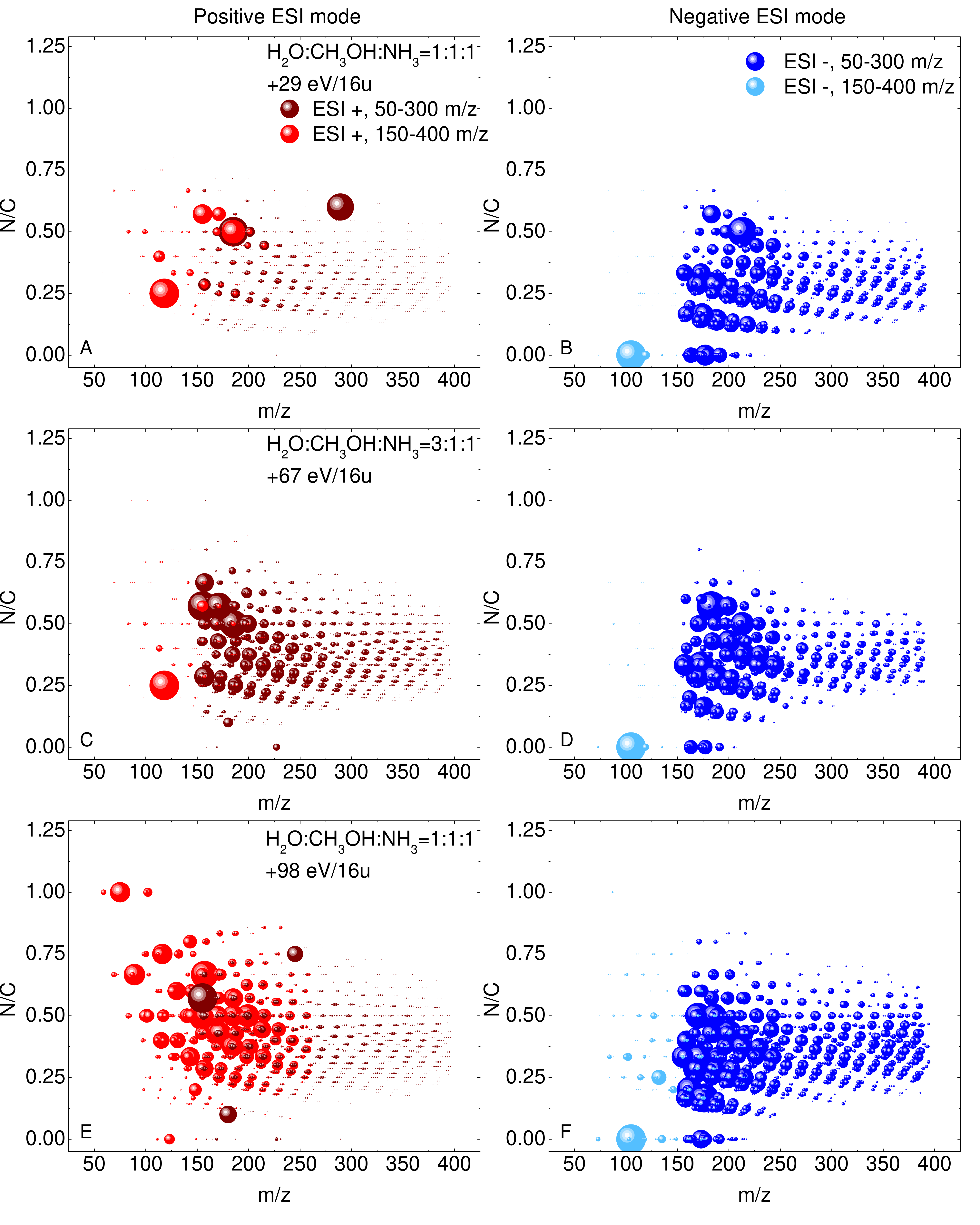}
		\caption{N/C ratios with respect to the m/z ratio in the range 50-300 and 150-400 m/z in positive ESI mode (left panels) and negative ESI mode (right panels). Panels A and B: 1:1:1+29 eV/16~u residue; panels C and D: 3:1:1+67 eV/16~u residue; panels E and F: residue of the 1:1:1+98 eV/16~u. The size of the dots is given by the normalized intensity.}
		\label{fig:NCMass}
	\end{figure*}

Figure~\ref{fig:NMass} and Fig.~\ref{fig:NCMass} and Fig.~\ref{fig:NCMass} show the elemental abundance of nitrogen as the number of N atoms and the N/C ratio with respect to the m/z. The interpretation of the different distribution of intensities in the ranges 50-300 and 150-400 m/z is given in sect.~\ref{VHRMS}.
	
	\section{HMT and its derivatives}
	
	\begin{figure*}
		\centering
		\includegraphics[width=0.8\linewidth]{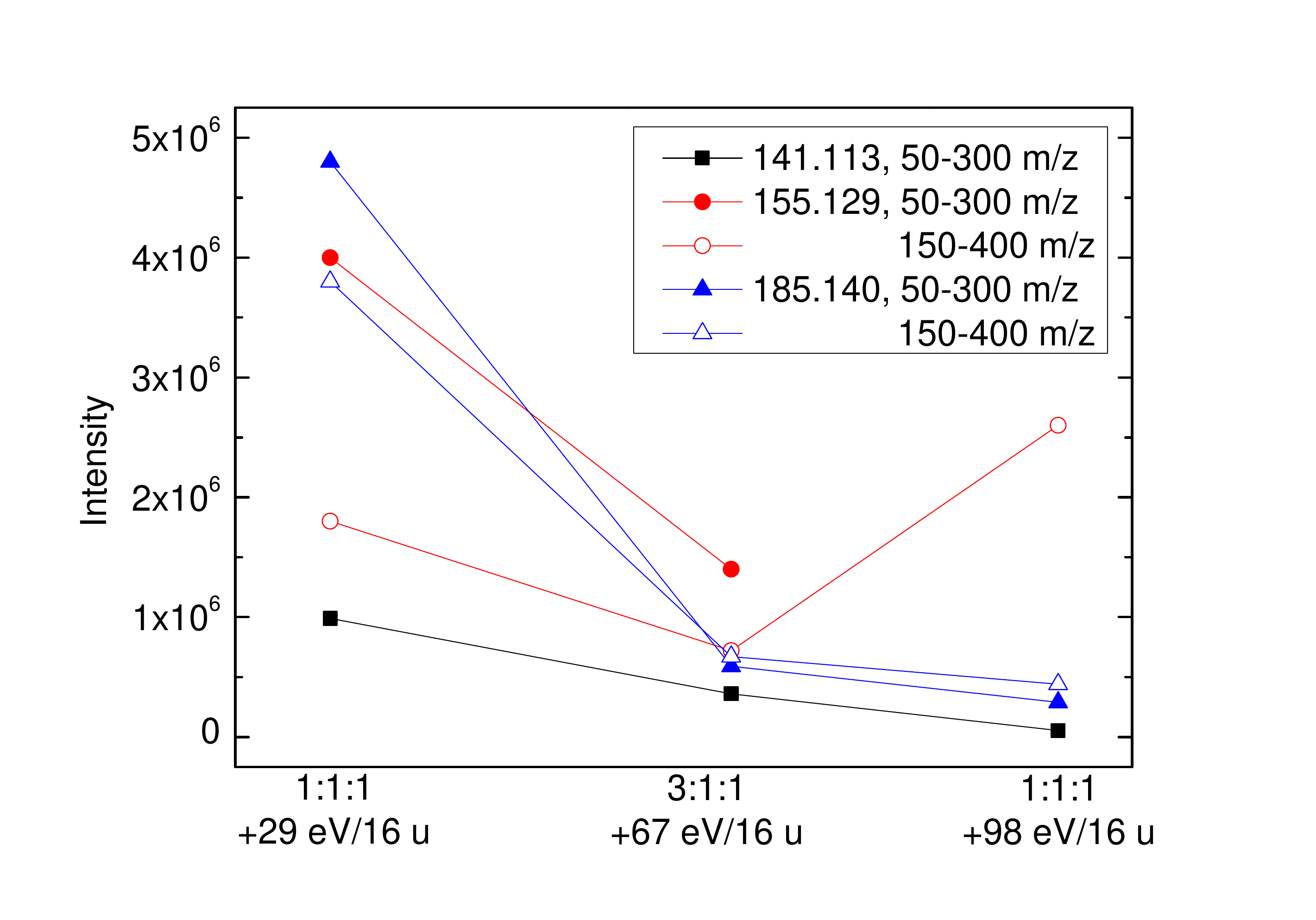}
		\caption{Intensity of m/z peaks of molecular ions with the same stoichiometric formulas of HMTH$^+$, HMT-CH$_3$H$^+$, and HMT-C$_2$H$_5$OH$^+$. }
		\label{fig:figurehmtabundance}
	\end{figure*}
	
	In Fig.~\ref{fig:figurehmtabundance} we show the intensity of peaks at 141.113, 155.129 and 185.140 m/z detected in the mass spectra of the residues in positive ESI modes in both the range 50-300 and 150-400 m/z. The intensity of both the 155.129 and 185.140 m/z peaks is higher than the intensity of the 141.113 m/z peak attributed to HMT. In VHRMS, the intensity of m/z peaks is associated to the abundance of the related species within the sample, implying that in ou HMT derivatives could be more abundant than HMT. 
	The higher intensity of the 155.129 m/z peak with respect to the 141.113 peak was also observed by \citet{danger13}, and it could be attributed to differences in the residue synthesis (mixture, dose, warm-up rate) or storage as well as to an artefact induced by the ionization method used in the VHRMS analysis.
	The intensity of the m/z peaks shown in Fig.~\ref{fig:figurehmtabundance} decrease according to the dose given to the pristine ice from which residues are produced. An anomaly is the high intensity of the 155.129 m/z peak detected in the 1:1:1+98 eV/16 u. In this sample, this molecular ion is only detected in the mass spectrum acquired between 150-400 m/z, and its intensity does not follow the trend observed for the 141.113 and 185.140 m/z peaks.
	
	\end{appendix}

\end{document}